\documentclass[journal]{IEEEtran}

\usepackage{amsmath,amssymb,amsfonts,mathrsfs}
\usepackage{braket}

\usepackage{graphicx}          
\usepackage{graphics} 
\usepackage{epsfig} 
\usepackage{times} 
\usepackage{floatrow}
\usepackage{enumerate}

\newtheorem{theorem}{Theorem}[section]
\newtheorem{corollary}{Corollary}[section]
\newtheorem{definition}{Definition}[section]
\newtheorem{lemma}{Lemma}[section]
\newtheorem{proposition}{Proposition}[section]
\newtheorem{remark}{Remark}[section]
\newtheorem{example}{Example}[section]

\def\bold{\boldsymbol}

\newcommand\Tstrut{\rule{0pt}{2.6ex}}         

\begin{document}

\title{Structural characterization of linear quantum systems with application to back-action evading measurement 
\thanks{
This research is supported in part by  a Hong Kong Research Grant council (RGC) grant (No. 15206915, No. 15208418), the Air Force Office of Scientific Research (AFOSR) and the office of Naval Research Grants (ONRS) under agreement number FA2386-16-1-4065, and  the Australian Research Council under grant number DP180101805. An extended archieve version can be found online (arXiv:1803.09419).}}
\author{Guofeng~Zhang, Ian~R.~Petersen,~\IEEEmembership{Fellow, IEEE}, and Jinghao Li \thanks{
G. Zhang is with the Department of Applied Mathematics, The Hong Kong
Polytechnic University, Hong Kong. (e-mail: Guofeng.Zhang@polyu.edu.hk).}
\thanks{
I.R. Petersen is with the Research School of Electrical, Energy and Material Engineering, The Australian National University, Canberra ACT 2601, Australia. (e-mail:
i.r.petersen@gmail.com).}
\thanks{J. Li is with College of Information Science and Engineering, Northeastern University, Shenyang 110819, China. (e-mail: lijinghao@ise.neu.edu.cn).}
}
\maketitle

\begin{abstract}

The purpose of this paper is to study the structure of quantum linear systems in terms of their Kalman canonical form, which was proposed in a recent paper \cite{ZGPG18}.  The spectral structure of quantum linear systems is explored, which indicates that a quantum linear system is both controllable and observable provided that it is Hurwitz stable.  A new parameterization method for quantum linear systems is proposed. This parameterization is designed for the Kalman canonical form directly. Consequently, the parameters involved are in a blockwise form in correspondence with the blockwise structure of the Kalman canonical form. This parameter structure can be used to simplify various quantum control design problems. For example,  necessary and sufficient conditions for the realization of quantum back-action evading (BAE) measurements  are given in terms of these new parameters. Due to their blockwise nature, a small number of parameters are required for realizing quantum BAE measurements.

\textbf{Index Terms---}  Quantum linear systems; Kalman canonical form; back-action evading (BAE) measurements.
\end{abstract}

\section{Introduction} \label{sec:intro}

The last two decades have witnessed a fast growth in the theoretical investigation and experimental demonstration  of quantum control as
it is an essential ingredient of quantum information technologies, including quantum communication, quantum computation, quantum cryptography, quantum ultra-precision metrology, and nano-electronics.  Quantum linear systems play an important role in quantum control theory. In the field of  quantum optics, linear systems are widely used as they are easy to manipulate and,
more importantly, they are often good approximations to more
general dynamics \cite{GZ00}, \cite{WM08}, \cite{WM10}, \cite{TNP+11}. Besides their wide
applications in quantum optical systems, quantum linear models have found important and successful
 applications for many other quantum dynamical systems  such as
opto-mechanical systems \cite{HRN+10}, \cite{TC12}, \cite{MTV+17}, circuit quantum electro-dynamical
(circuit QED) systems \cite{MJP+11}, and atomic ensembles \cite{SvHM04}, \cite{MTV+17}.

In \cite{ZGPG18}, a Kalman canonical form is proposed for quantum linear
systems. More specifically, given a quantum linear system, an orthogonal and
blockwise symplectic transformation matrix is constructed which transforms the original system
into four subsystems: the controllable and observable ($co$) subsystem, the
controllable and unobservable ($c\bar{o}$) subsystem, the uncontrollable and
observable ($\bar{c}o$) subsystem, and the uncontrollable and unobservable ($
\bar{c}\bar{o}$) subsystem. In \cite{ZGPG18}, the quantities $\bold{x}_{co}, \bold{x}_{\bar{c}\bar{o}}, \bold{q}_h,\bold{p}_h$ are used to denote the quadrature operators of the $co, \bar{c}\bar{o}, c\bar{o}, \bar{c}o$ subsystems respectively. The Kaman canonical form has also been derived in Reference \cite{GZPZ17} by means of an  SVD-like factorization.

On the basis of the Kalman canonical form proposed in \cite{ZGPG18}, in this paper we aim to explore deeper the structure of quantum linear systems. An open quantum system can be described by a set of quantum stochastic differential equations (QSDEs). Due to the nature of quantum-mechanical systems, there are constraints on the coefficients of these QSDEs, which are called physical realizability conditions of  quantum systems \cite{JNP08}. In this paper we present physical realizability conditions for quantum linear systems in  their Kalman canonical form; see Lemma \ref{cor:ABC}. Interestingly, these conditions allow us to expose the nice structure of the spectrum of a quantum linear system; see Propositions \ref{cor:barco} and \ref{rem:h:eig}, and Example \ref{ex:h}. Moreover, it is shown in Theorem \ref{thm:hurwitz} that if a quantum linear system is Hurwitz stable,, then it is both controllable and observable.  A new parameterization method for quantum linear systems is proposed in the paper.  We express the system Hamiltonian and the coupling operator explicitly in terms of the  partitioned system variables $\bold{x}_{co}, \bold{x}_{\bar{c}\bar{o}}, \bold{q}_h,\bold{p}_h$.  Specifically, let $\bold{x} = [\bold{q}_h^\top \ \bold{p}_h^\top \ \bold{x}_{co}^\top \  \bold{x}_{\bar{c}\bar{o}}^\top]^\top$. Then the  system Hamiltonian is $\bold{H} = \bold{x}^\top H\bold{x}/2$ where $H$ is a real symmetric matrix, and the coupling operator is $\bold{L} = \Gamma\bold{x}$ with $\Gamma$ being a complex matrix.   Due to the special structure of the system matrices in the Kalman canonical form, if $\bold{H}$ and $\bold{L}$ generate the Kalman canonical form, then the matrices $H$ and $\Gamma$ should be of specific form. This form  is given in Lemma \ref{lem:H_Gamma}. Moreover, we also establish the converse: If the matrices $H$ and $\Gamma$ are of  the given specific form, then the resulting QSDEs are formally in  the Kalman canonical form; see Lemma \ref{thm:ABC_H_Gamma}. Finally, as the Kalman canonical form is obtained based on the notions of controllability and observability, we derive further conditions on $H$ and $\Gamma$ such that the resulting system is indeed the quantum Kalman canonical form;  see Theorem \ref{thm:co}.

A measurement process often involves measurement noise from the  surrounding
environment. In quantum mechanics, environmental noise can be represented by
two conjugate quadrature operators. A fundamental fact in quantum
mechanics is that these two noise quadrature operators do not commute. This
gives rise to the so-called standard quantum limit (SQL). However, if a
measurement process suffers from a noise quadrature (shot noise), but not from
the conjugate quadrature noise (measurement back-action noise), then it is
called a BAE measurement, \cite{CVB+07}, \cite[Fig.\;2(a)]{TC10}, \cite{WC13},   \cite{NY14}, \cite{ZZZ+15}, \cite{ODP+16}, \cite{YY16}.  As a result, a BAE
measurement may be able to beat the SQL, thus enabling extremely high precision
measurement. In fact, the idea of BAE measurement originates from the study
of gravitational wave  detection \cite{HRN+10}. In the language of
linear systems theory, a BAE measurement is realized if
the transfer function from the measurement back-action noise to the measured
output is zero.  On the basis of the proposed new parameterization method for quantum linear systems, in this paper, necessary and sufficient conditions for the realization of BAE measurements   by means of quantum linear systems are given in Theorem \ref{thm:BAE}.

The rest of the paper is organized as follows. The notation commonly used in this paper is summarized in Subsection \ref{subsec:notation}.  Preliminaries are given in Section \ref {sec:pre}, which include quantum linear systems and their Kalman canonical form.  The structural properties of the Kalman canonical form are studied in Section \ref{sec:Kalman}. An application to the realization of quantum BAE measurements is studied in Section \ref{sec:application}. Examples are given in Section \ref{sec:example}. Concluding remarks are given in Section \ref{sec:con}.

\vspace{-2.3mm}
\subsection{Notation}\label{subsec:notation}

\begin{enumerate}
\item $x^{\ast }$ denotes the complex conjugate of a complex number $x$ or
the adjoint of an operator $x$. 

\item For a matrix $X=[x_{ij}]$ with number or operator entries,  denote $
X^{\#}=[x_{ij}^{\ast }]$, $X^{\top}=[x_{ji}]$, and $
X^{\dag }=(X^{\#})^{\top }$. Moreover, let $\breve{X}=\bigl[
\begin{smallmatrix}
X \\
X^{\#}
\end{smallmatrix}
\bigr]$. $\mathrm{Re}(X) $ and $\mathrm{Im}(X) $  denote the real part and imaginary part of a matrix $X$, respectively.

\item The commutator of two operators $X$ and $Y$
is defined as $[X,Y]\triangleq XY-YX$. If $X$ and $Y$ are two vectors of self-adjoint
operators, then their commutator is defined as the matrix of operators $[
X,Y^{\top}] \triangleq XY^{\top}-(YX^{\top})^{\top}$.

\item $I_{k}$ is the identity matrix and $0_{k}$ the zero matrix in $
\mathbb{C}^{k\times k}$. $\delta _{ij}$ denotes the Kronecker delta.  Let $J_k = \mathrm{diag}(I_{k},-I_{k})$. For a matrix $X\in
\mathbb{C}^{2k\times 2r}$, define its $\flat $-adjoint by $X^{\flat
}\triangleq J_{r}X^{\dag }J_{k}$.

\item Given two matrices $U$, $V\in \mathbb{C}^{k\times r}$, define $\Delta
(U,V)\triangleq \bigl[
\begin{smallmatrix}
U& V \\
U^\# & V^\#
\end{smallmatrix}
\bigr]$. A matrix with this structure will be called \emph{doubled-up} \cite{GJN10}.

\item A matrix $T\in \mathbb{C}^{2k\times 2k}$ is called \emph{Bogoliubov}
if it is doubled-up and satisfies $TT^{\flat }=T^{\flat }T=I_{2k}$.

\item Let $\mathbb{J}_{k} = \bigl[
\begin{smallmatrix}
0_{k} & I_{k} \\
-I_{k} & 0_{k}
\end{smallmatrix}
\bigr]$.  A matrix $S\in \mathbb{C}^{2k\times 2k}$ is called \emph{symplectic},
if it satisfies $S\mathbb{J}_{k}S^{\dag }=S^{\dag }\mathbb{J}_{k}S=\mathbb{J}_{k}$.
\end{enumerate}

\section{Quantum linear systems and their Kalman canonical form} \label{sec:pre}
In this section, quantum linear systems are briefly introduced, and their Kalman canonical form, recently derived in \cite{ZGPG18},  is also presented for completeness.

\subsection{Quantum linear systems} \label{sec:qls}

\begin{figure}[tbph]
\centering
\includegraphics[width=0.6\textwidth]{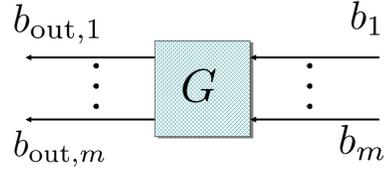}
\caption{Open quantum linear system $G$ composed of $n$ harmonic oscillators driven by $m$ input fields.}
\label{fig:sys}
\end{figure}

The open quantum linear system $G$, as shown in Fig. \ref{fig:sys}, can be used to model a collection
of $n$ quantum harmonic oscillators interacting with $m$ input boson fields. The $j$-th
 oscillator, $j=1,\ldots ,n$, may be represented by its
annihilation operator $\boldsymbol{a}_{j}$ and creation operator $
\boldsymbol{a}_{j}^{\ast }$ (the adjoint operator of $\boldsymbol{a}_{j}$).
These are operators on an infinite-dimensional Hilbert space and satisfy the \emph{canonical
commutation relations (CCRs)}
$[\boldsymbol{a}_{j}(t), \boldsymbol{a}_{k}(t)]=[\boldsymbol{a}_{j}^{\ast }(t), \boldsymbol{a}_{k}^{\ast }(t)]=0$, and
$[\boldsymbol{a}_{j}(t), \boldsymbol{a}_{k}^{\ast }(t)]=\delta _{jk}$,
 $\forall j,k=1,\ldots n,\forall t\in \mathbb{R}^{+}$. Let $
\boldsymbol{a}=[\boldsymbol{a}_{1}~\cdots ~\boldsymbol{a}_{n}]^{\top }$. The
system Hamiltonian is given by
$\boldsymbol{H}=(1/2)\boldsymbol{\breve{a}}^{\dag }\Omega \boldsymbol{\breve{a}}$, where
$\boldsymbol{\breve{a}}
=
[\boldsymbol{a}^{\top } \; (\boldsymbol{a}^{\#})^{\top}]^{\top }$, and
$\Omega =\Delta (\Omega _{-},\Omega _{+})\in \mathbb{C}^{2n\times 2n}$
is a Hermitian matrix with $\Omega _{-},\Omega _{+}\in
\mathbb{C}^{n\times n}$. The coupling of the system to the input fields is
described by the operator
$\boldsymbol{L}=[C_{-} \ C_{+}]\boldsymbol{\breve{a}}$,
with $C_{-},C_{+}\in \mathbb{C}^{m\times n}$. The $k$-th input boson field,
$k=1,\ldots ,m$, is represented in terms of its annihilation operator
$\boldsymbol{b}_{k}(t)$ and  creation operator $\boldsymbol{b}_{k}^{\ast}(t)$ (the adjoint operator of
$\boldsymbol{b}_{k}(t)$). These are operators on a symmetric Fock space (a special kind of infinite-dimensional Hilbert space, \cite{P92}). The operators $\boldsymbol{b}_{k}(t)$ and
$\boldsymbol{b}_{k}^{\ast}(t)$ satisfy the {\it singular commutation relations}
$[\boldsymbol{b}_{j}(t), \boldsymbol{b}_{k}(r)]=[\boldsymbol{b}_{j}^{\ast }(t), \boldsymbol{b}_{k}^{\ast }(r)]=0$, and
$[\boldsymbol{b}_{j}(t),\ \boldsymbol{b}_{k}^{\ast}(r)]=\delta _{jk}\delta (t-r)$,
$\forall j,k=1,\ldots ,m,~\forall t,r\in \mathbb{R}$. Let
$\boldsymbol{b}(t)=[\boldsymbol{b}_{1}(t)\ \cdots ~\boldsymbol{b}_{m}(t)]^{\top }$ and
$\boldsymbol{\breve{b}}(t)=[\boldsymbol{b}(t)^{\top }\ (\boldsymbol{b}(t)^{\#})^{\top }]^{\top }$.

The dynamics of the open quantum linear system in Fig. \ref{fig:sys} is
described by the following QSDEs, (\cite[Eq. (26)]{GJN10}, \cite[Eqs. (14)-(15)]{ZJ13})
\begin{equation}\label{eq:sys_a}
\begin{split}
\boldsymbol{\dot{\breve{a}}}(t)
=&\; \mathcal{A}\boldsymbol{\breve{a}}(t)+\mathcal{B}\boldsymbol{\breve{b}}(t),
 \\
\boldsymbol{\breve{b}}_{\mathrm{out}}(t)
=&\; \mathcal{C}\boldsymbol{\breve{a}}(t)+\boldsymbol{\breve{b}}(t),
\ \ t\geq 0,
\end{split}
\end{equation}
where the system matrices are  parametrized by the physical parameters of the Hamiltonian $\bold{H}$ an coupling $\bold{L}$, which are
\begin{equation*}\label{ABC}
\mathcal{C}=\Delta (C_{-},\ C_{+}), ~\mathcal{B}=-\mathcal{C}^{\flat }, ~   \mathcal{A}=-\imath J_{n}\Omega -\frac{1}{2}\mathcal{C}^{\flat }\mathcal{C}.
\end{equation*}
These system matrices satisfy
\begin{equation} \label{eq:PR_a}
\mathcal{A}+\mathcal{A}^{\flat }+\mathcal{B}\mathcal{B}^{\flat } = 0, \ \
\mathcal{B} =-\mathcal{C}^{\flat }.
\end{equation}
On the other hand, if matrices $\mathcal{A},\mathcal{B},\mathcal{C}$ satisfy Eq. \eqref{eq:PR_a},  then the system Hamiltonian $\boldsymbol{H}=(1/2)\boldsymbol{\breve{a}}^{\dag }\Omega \boldsymbol{\breve{a}}$ is completely determined as the matrix $\Omega$ can be computed via
\begin{equation}\label{Omega}
\Omega =\frac{\imath}{2}( J_{n}\mathcal{A-A}^{\dag }J_{n}) .
\end{equation}
Moreover, the coupling $\boldsymbol{L}=[C_{-} \ C_{+}]\boldsymbol{\breve{a}}$ is also determined once the matrix $\mathcal{C}$ is given. In this case, the mathematical model \eqref{eq:sys_a}  is said to be \textit{physically realizable}  as it  could in principle be physically realized (\cite{JNP08}, \cite{NJD09}, \cite{IRP11}).


\subsection{The quantum Kalman canonical form} \label{sec:zgpg18}

The Kalman decomposition of quantum linear systems,  recently developed in  \cite{ZGPG18}, is presented in this subsection.

A unitary and blockwise Bogoliubov coordinate transformation matrix $T$ is defined in
\cite[Eq. (47)]{ZGPG18}, which is
\begin{equation*}
T
\triangleq 
\left[
\begin{array}{cc|cc|cc}
Z_{3} & 0 & Z_{1} & 0 & Z_{2} & 0 \\
0 & Z_{3}^{\#} & 0 & Z_{1}^{\#} & 0 & Z_{2}^{\#}
\end{array}
\right] ,
\label{T}
\end{equation*}
where $Z_1 \in \mathbb{C}^{n\times n_1}$, $Z_2 \in \mathbb{C}^{n\times n_2}$, and
 $Z_3 \in \mathbb{C}^{n\times n_3}$ \ ($n_{1},n_{2},n_{3}\geq 0$ and
 $n_{1}+n_{2}+n_{3}=n$).
 $T$ is called blockwise Bogoliubov as it satisfies
\begin{equation}  \label{TJT}
T^{\dag }J_{n}T
=
\left[
\begin{array}{ccc}
J_{n_{3}} & 0 & 0 \\
0 & J_{n_{1}} & 0 \\
0 & 0 & J_{n_{2}}
\end{array}
\right] .
\end{equation}
Define the unitary matrix $\Pi \in \mathbb{C}^{2n_{3}\times 2n_{3}}$ by
\begin{equation*}
\Pi \triangleq
\left[
\begin{array}{cccc}
I_{n_{a}} & 0 & 0 & 0 \\
0 & 0 & 0 & -I_{n_{b}} \\
0 & 0 & I_{n_{a}} & 0 \\
0 & I_{n_{b}} & 0 & 0
\end{array}
\right] ,
\end{equation*}
where $0\leq n_{a},n_{b}\leq n_{3}$, and $n_{a}+n_{b}=n_{3}$. Let $\tilde{V}_{n_{3}}= \Pi V_{n_{3}}$,
where
\begin{equation*} \label{V_k}
V_{k} \triangleq \frac{1}{\sqrt{2}}\left[
\begin{array}{cc}
I_{k} & I_{k} \\
-\imath I_{k} & \imath I_{k}
\end{array}
\right] ,\ \ k\in \mathbb{N}
\end{equation*}
is a unitary matrix. Define two more unitary matrices
\begin{equation*}
\tilde{V}_{n}\triangleq \left[
\begin{array}{ccc}
\tilde{V}_{n_{3}} &  & \parbox{12pt}{\Huge 0} \\
& V_{n_{1}} &  \\
\parbox{12pt}{\Huge 0} &  & V_{n_{2}}
\end{array}
\right]
\end{equation*}
and
\begin{equation*} \label{hatT}
\hat{T}\triangleq T\tilde{V}_{n}^{\dag }.
\end{equation*}

The following result is proved in \cite{ZGPG18}, which puts the quantum linear  system (\ref{eq:sys_a}) in the Kalman
canonical form.

\begin{figure}
  \centering
  \includegraphics[width=0.7\textwidth]{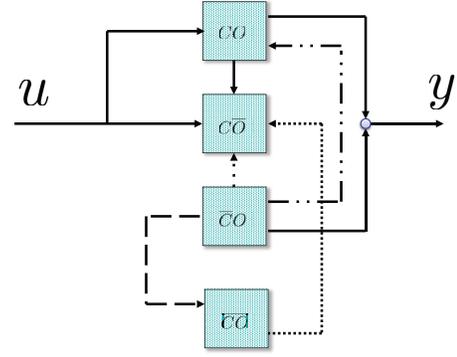}\\
  \caption{The Kalman canonical form of a quantum linear system; see \cite[Fig.~2]{ZGPG18}.} \label{KD}
\end{figure}

\begin{proposition}\cite[Theorem 4.4]{ZGPG18}\label{thm:general_Kalman_4}
 The coordinate transformations
\begin{equation} \label{complex-real}
\begin{split}
&\left[
\begin{array}{c}
\boldsymbol{q}_{h} \\
\boldsymbol{p}_{h} \\\hline
\boldsymbol{x}_{co} \\\hline
\boldsymbol{x}_{\bar{c}\bar{o}}
\end{array}
\right]\equiv \boldsymbol{x} \triangleq \hat{T}^{\dag }\boldsymbol{\breve{a}},
\\
&\left[
\begin{array}{c}
\boldsymbol{q}_{\mathrm{in}} \\
\boldsymbol{p}_{\mathrm{in}}
\end{array}
\right] \equiv \boldsymbol{u}\triangleq V_{m}\boldsymbol{\breve{b}},\;
\left[
\begin{array}{c}
\boldsymbol{q}_{\mathrm{out}} \\
\boldsymbol{p}_{\mathrm{out}}
\end{array}
\right] \equiv \boldsymbol{y}\triangleq V_{m}\boldsymbol{\breve{b}}_{\mathrm{
out}}
\end{split}
\end{equation}
convert the system (\ref{eq:sys_a}) into the real quadrature form
\begin{equation} \label{real_Kalman_ss}
\begin{split}
\boldsymbol{\dot{x}}(t) 
=&\;
\bar{A}\boldsymbol{x}(t)  +\bar{B}\boldsymbol{u}(t),
\\
\boldsymbol{y}(t)
=&\;
 \bar{C}\boldsymbol{x}(t) +\boldsymbol{u}(t),
\end{split}
\end{equation}
where the real matrices $\bar{A},\bar{B},\bar{C}$ are of the form
\begin{equation}  \label{real_Kalman_sys_A}
\begin{split}
\bar{A}
\triangleq &
\left[
\begin{array}{cc|c|c}
A_{h}^{11} & A_{h}^{12} & A_{12} & A_{13} \\
0 & A_{h}^{22} & 0 & 0 \Tstrut \\ \hline
0 & A_{21} & A_{co} & 0 \Tstrut  \\ \hline
0 & A_{31} & 0 & A_{\bar{c}\bar{o}} \Tstrut
\end{array}
\right] ,
\\
\bar{B}
\triangleq &
\left[
\begin{array}{c}
B_{h} \\
0 \\ \hline
B_{co} \\ \hline
0
\end{array}
\right] ,   \   \
\bar{C}
\triangleq
\left[
\begin{array}{cc|c|c}
0 & C_{h} & C_{co} & 0
\end{array}
\right].
\end{split}
\end{equation}
The corresponding  system block diagram is shown in Fig.~\ref{KD}.
\end{proposition}

\section{Structure of the quantum Kalman canonical form} \label{sec:Kalman}
In this section, we investigate the structure of the quantum Kalman canonical form. A useful lemma is given in Subsection \ref{sec:lemmas}, the spectral structure of  the quantum Kalman canonical form is presented in Subsection \ref{sec:spectral}, a new parametrization method for quantum Kalman canonical form is proposed in Subsection \ref{sec:parameterization}, and finally a refined form of the quantum Kalman canonical form is presented in Subsection \ref{subsec:app1}.

\subsection{A useful lemma} \label{sec:lemmas}
In this subsection, we present a lemma, Lemma \ref{cor:ABC}, which is useful for the future development of this section.

The following result is an immediate consequence of the coordinate transformations \eqref{complex-real} and the identity $\hat{T}^{\dag }J_{n}\hat{T}  =  \imath \mathbb{\bar{J}}_{n}$, where
\[
\mathbb{\bar{J}}_{n}
\triangleq
\left[
\begin{array}{ccc}
\mathbb{J}_{n_{3}} & 0 & 0 \\
0 & \mathbb{J}_{n_{1}} & 0 \\
0 & 0 & \mathbb{J}_{n_{2}}
\end{array}
\right].
\]

\begin{proposition}\label{lem:pr}
For the Kalman canonical form (\ref{real_Kalman_ss}), the following  conditions hold.
\begin{equation} \label{pr_1a}
\begin{split}
\bar{A}\mathbb{\bar{J}}_{n}+\mathbb{\bar{J}}_{n}\bar{A}^{\top}+\bar{B}\mathbb{J}
_{m}\bar{B}^{\top}
=&\;
0,
\\
\bar{B}
=&\;
\mathbb{\bar{J}}_{n}\bar{C}^{\top}\mathbb{J}_{m}. 
\end{split}
\end{equation}
\end{proposition}


Substituting system matrices $\bar{A}, \bar{B}$ and $\bar{C}$ in Eq. (\ref{real_Kalman_sys_A}) into Eq. (\ref{pr_1a}) we can obtain the following result which presents the real-quadrature counterpart of physical realizability conditions \eqref{eq:PR_a}.   The proof is  straightforward, hence it is omitted.

\begin{lemma}\label{cor:ABC}
 For the Kalman canonical form (\ref{real_Kalman_ss}), the following conditions hold.
\begin{subequations}
\begin{align}
A_{h}^{22^{\top}}
=&\;
-A_{h}^{11},
\label{pr1a_a}
 \\
-A_{h}^{12}+A_{h}^{12^{\top}}+B_{h}\mathbb{J}_{m}B_{h}^{\top}
=&\;
 0,
 \nonumber
\\
-A_{12}+A_{21}^{\top}\mathbb{J}_{n_{1}}+B_{h}\mathbb{J}_{m}B_{co}^{\top}\mathbb{J}_{n_{1}}
=&\;
 0,
  \nonumber
  \\
A_{31}^{\top}\mathbb{J}_{n_{2}}
=&\;
 A_{13},
   \nonumber
 \\
\mathbb{J}_{n_{1}}A_{co}+A_{co}^{\top}\mathbb{J}_{n_{1}}
-\mathbb{J}_{n_{1}}B_{co}\mathbb{J}_{m}B_{co}^{\top}\mathbb{J}_{n_{1}}
=&\;
 0,
   \nonumber
\\
\mathbb{J}_{n_{2}}A_{\bar{c}\bar{o}}+A_{\bar{c}\bar{o}}^{\top}\mathbb{J}_{n_{2}}
=&\;
 0,
 \label{pr1a_g}
\end{align}
and
\begin{align}
B_{h}
=&\;
C_{h}^{\top}\mathbb{J}_{m}, \ 
B_{co}
=
\mathbb{J}_{n_{1}}C_{co}^{\top}\mathbb{J}_{m}.
\label{pr1b_b}
\end{align}
\end{subequations}
\end{lemma}

\subsection{Spectral structure of the quantum Kalman canonical form} \label{sec:spectral}

The specific relations among the components of the system matrices $\bar{A}, \bar{B}$ and $\bar{C}$, established in Lemma \ref{cor:ABC}, can be used to explore the spectral structure of quantum linear systems, which is the focus of this subsection.  We denote the set of eigenvalues of a matrix $A$ by $\sigma(A)$.

Let us first look at the $\bar{c}\bar{o}$ subsystem by ignoring the other
modes in  the Kalman canonical form \eqref{real_Kalman_ss},  which is
\begin{equation}\label{barco_system}
\boldsymbol{\dot{x}}_{\bar{c}\bar{o}}(t)
=
A_{\bar{c}\bar{o}}\boldsymbol{x}_{\bar{c}\bar{o}}(t).
\end{equation}
The following result  shows that the poles of this subsystem are symmetric about both the real and imaginary axes.

\begin{proposition} \label{cor:barco}
If $\lambda \in \sigma (A_{\bar{c}\bar{o}})$, then
$-\lambda ,\lambda ^{\ast},-\lambda ^{\ast }\in \sigma (A_{\bar{c}\bar{o}})$.
\end{proposition}

{\it Proof.} Let $\lambda $ and $\mu $ be an eigenvalue and eigenvector of the matrix $A_{\bar{c}\bar{o}}$; i.e.,
$A_{\bar{c}\bar{o}}\mu =\lambda \mu $. By Eq. (\ref{pr1a_g}),
 $-\lambda \mathbb{J}_{n_{2}}\mu
 =-\mathbb{J}_{n_{2}}A_{\bar{c}\bar{o}}\mu
=A_{\bar{c}\bar{o}}^{\top}\mathbb{J}_{n_{2}}\mu $.
 In other words, $-\lambda $
is an eigenvalue of $A_{\bar{c}\bar{o}}^{\top}$ with the corresponding
eigenvector $\mathbb{J}_{n_{2}}\mu $. As
$\sigma (A_{\bar{c}\bar{o}})=\sigma(A_{\bar{c}\bar{o}}^{\top})$, we have
 $-\lambda \in \sigma (A_{\bar{c}\bar{o}})$. Therefore, if $\lambda \in \sigma (A_{\bar{c}\bar{o}})$, then
 $-\lambda,\lambda ^{\ast },-\lambda ^{\ast }\in \sigma (A_{\bar{c}\bar{o}})$. \hfill $\Box $

In  the Kalman canonical form (\ref{real_Kalman_ss}),
if we ignore the $\bar{c}\bar{o}$ and $co$ subsystems, we obtain the
following subsystem
\begin{equation}\label{sys_h1}
\begin{split}
\left[
\begin{array}{c}
\boldsymbol{\dot{q}}_{h}(t) \\
\boldsymbol{\dot{p}}_{h}(t)
\end{array}
\right]
=&\;
\left[
\begin{array}{cc}
A_{h}^{11} & A_{h}^{12} \\
0 & A_{h}^{22} \Tstrut
\end{array}
\right] \left[
\begin{array}{c}
\boldsymbol{q}_{h}(t) \\
\boldsymbol{p}_{h}(t)
\end{array}
\right]
+\left[
\begin{array}{c}
B_{h} \\
0
\end{array}
\right] \boldsymbol{u}(t),
 \\
\boldsymbol{y}(t)
=&\;
\left[
\begin{array}{cc}
0 & C_{h}
\end{array}
\right] \left[
\begin{array}{c}
\boldsymbol{q}_{h}(t) \\
\boldsymbol{p}_{h}(t)
\end{array}
\right] +\boldsymbol{u}(t).
\end{split}
\end{equation}
In this paper,  the system (\ref{sys_h1}) is called the ``h''
subsystem.

\begin{proposition}\label{rem:h:eig}
For the ``h'' subsystem (\ref{sys_h1}), we have:
\begin{description}
\item[1)] The set of the poles is given by $
\sigma \left( A_{h}^{11}\right) \cup \sigma \left( -A_{h}^{11}\right)$;

\item[2)] If $\lambda $ is a pole of this subsystem, then so are $-\lambda ,\lambda ^{\ast },-\lambda
^{\ast }$.
\end{description}
\end{proposition}

{\it Proof.} Item 1) is an immediate
consequence of Eq. (\ref{pr1a_a}), while Item 2) follows from Item 1). \hfill $\Box$

If $C_+=0$ and $\Omega_+=0$, the resulting quantum linear system \eqref{eq:sys_a} is said to
be {\it passive} (\cite{ZJ11}, \cite{ZJ13}, \cite{NY14}, \newline \cite{GZ15}, \cite{GY13}). In the passive case, the existence of purely imaginary poles is
equivalent to the existence of the $\bar{c}\bar{o}$ subsystem, as has been
proved in \cite[Theorem 3.2]{ZGPG18}. In the general (not necessarily passive) case, according to  Propositions \ref{cor:barco}-\ref{rem:h:eig}, the poles of the ``h'' subsystem and $\bar{c}\bar{o}$ subsystem  are symmetric about the real and imaginary axes. However, this spectral property does not
guarantee the existence of an ``h'' subsystem or  a $\bar{c}\bar{o}$ subsystem. This is shown by the following
counter-example.

\begin{example} \label{ex:h}
Let
\begin{equation*}
A =\left[
\begin{array}{cc}
1 & 0 \\
0 & -1
\end{array}
\right] ,\ \ B=\left[
\begin{array}{cc}
1 & 0\\
2&0
\end{array}
\right],  \ \
C=\left[
\begin{array}{cc}
0 &0 \\
2&-1
\end{array}
\right] .
\end{equation*}
It is easy to see that the two poles of the system $\left( A,B, C\right) $ are $-1$ and $1$, which are symmetric about the real and imaginary axes. Moreover, this system is physically realizable as it satisfies Eq. (\ref{pr_1a}).  However, this system is both controllable and observable. Thus it is neither an ``h'' subsystem nor a  $\bar{c}\bar{o}$ subsystem.
\end{example}

Finally, we look at the $co$ subsystem by ignoring the other modes in the Kalman canonical form  \eqref{real_Kalman_ss}, which is
\begin{equation}\label{sys_co1_feb18}
\begin{split}
\boldsymbol{\dot{x}}_{co}(t)
=&\;
A_{co}\boldsymbol{x}_{co}(t)+B_{co}\boldsymbol{u}(t),
\\
\boldsymbol{y}(t)
=&\;
 C_{co}\boldsymbol{x}_{co}(t)+\boldsymbol{u}(t).
\end{split}
\end{equation}
In general, the poles of a $co$ subsystem are not symmetric about the real and imaginary axes.  For example, let $n=m=1$, $\Omega _{-}=C_{+}=0$, and $C_{-}=\Omega _{+}=1$. It
is easy to see that the resulting quantum linear system is both controllable
and observable; in other words, it is a $co$ system. However, the poles of this system are $1/2$ and $-3/2$, which are not symmetric about the real and imaginary axes.

The spectral structure of the Kalman canonical form established above implies the following interesting result.

\begin{theorem}\label{thm:hurwitz}
If a quantum linear system is Hurwitz stable, then it is both controllable and observable.
\end{theorem}

{\it Proof.} Assume a given quantum linear system is Hurwitz stable;  i.e., all its poles are on the open left-half plane. Without loss of generality, suppose the system is in the Kalman canonical form \eqref{real_Kalman_ss}.  According to the form of the matrix $\bar{A}$ in Eq. \eqref{real_Kalman_sys_A}, or equivalently \cite[Eq. (74)]{ZGPG18}, the poles of the system are those of the $\bar{c}\bar{o}$, $co$,  and ``h'' subsystems.   However, by Propositions \ref{cor:barco} and \ref{rem:h:eig},  there must be no the $\bar{c}\bar{o}$ and ``h'' subsystems. In other words, in this case, the only subsystem in the Kalman canonical form \eqref{real_Kalman_ss} is the $co$ subsystem. This means that the quantum linear system is  both controllable and observable. \hfill $\Box$

Theorem \ref{thm:hurwitz}  tells us that Hurwitz stability implies controllability and observability for quantum linear systems; in general the converse is not true, as  shown by Example \ref{ex:h} above. However, for the passive case, Theorem \ref{thm:hurwitz} can be strengthened to the following result, which  has already been proved in \cite[Lemma 2]{GZ15}.

\begin{corollary}\label{cor:hurwitz}
For a passive quantum linear system, the properties of  Hurwitz stability, controllability and observability are all equivalent.
\end{corollary}


We end this section with a final remark.

\begin{remark}\label{rem:Hamiltonian}
A $2d\times 2d$ real matrix $N$ is said to be a Hamiltonian matrix if the matrix $\mathbb{J}_d N$ is symmetric; see, e.g., \cite[Fact 3.19.1]{Ber09}.  If a Hamiltonian matrix has $\lambda$  as an eigenvalue, then  $-\lambda, \lambda ^{\ast }, -\lambda^{\ast }$ are also its eigenvalues. Later in Lemma \ref{thm:ABC_H_Gamma} and Remark \ref{rem:Hamiltonian2} we will show that both the matrices $A_{\bar{c}\bar{o}}$ and
 $\begin{smallmatrix}\left[
\begin{array}{cc}
A_{h}^{11} & 0 \\
0 & A_{h}^{22}
\end{array}
\right]
\end{smallmatrix}$  are Hamiltonian matrices, while in general the matrix $A_{co}$ is not.
\end{remark}

\subsection{Parameterization for the quantum Kalman canonical form}\label{sec:parameterization}
For the quantum Kalman canonical form (\ref{real_Kalman_ss}), let the system Hamiltonian  be
\begin{equation} \label{H}
\boldsymbol{H}
=
\frac{1}{2}\boldsymbol{x}^{\top}H\boldsymbol{x},
\end{equation}
where the real-quadrature operator $\boldsymbol{x}$ satisfies the CCRs $[\boldsymbol{x}(0), \ \boldsymbol{x}(0)^\top]=\imath \mathbb{J}_n$, and  the real matrix $H\in \mathbb{R}^{2n\times 2n}$ is symmetric.
Let the coupling operator be
\begin{equation}\label{L}
\boldsymbol{L}=\Gamma\boldsymbol{x},
\end{equation}
where $\Gamma\in \mathbb{C}^{m\times 2n}$.  In this subsection, we aim to find conditions on the matrices $H$ and $\Gamma$ such that the QSDEs generated by  the system Hamiltonian $\boldsymbol{H}$ in Eq. (\ref{H}) and the coupling operator $\boldsymbol{L}$ in Eq. (\ref{L}) are exactly the Kalman canonical form (\ref{real_Kalman_ss}).

\subsubsection{The necessary condition}\label{subsec:necessary}

By means of Lemma \ref{cor:ABC} given in subsection \ref{sec:lemmas}, we can
derive the following result, which presents a necessary condition for the system Hamiltonian
$\boldsymbol{H}$ in Eq. (\ref{H}) and the coupling operator $\boldsymbol{L}$ in Eq. (\ref{L}) to generate the quantum Kalman canonical form (\ref{real_Kalman_ss}).

\begin{lemma}\label{lem:H_Gamma}
If the system Hamiltonian $\boldsymbol{H}$ in Eq. (\ref{H}) and the coupling operator $\boldsymbol{L}$ in Eq. (\ref{L}) lead to QSDEs in  the Kalman canonical form (\ref{real_Kalman_ss}), then the real
symmetric matrix $H$ must be of the form
\begin{equation}\label{R}
H=\left[
\begin{array}{cc|c|c}
0 & H_{h}^{12} & 0 & 0 \\
H_{h}^{12^{\top}} & H_{h}^{22} & H_{12} & H_{13}
 \\ \hline
0 & H_{12}^{\top} & H_{co} & 0 \Tstrut
 \\ \hline
0 & H_{13}^{\top} & 0 & H_{\bar{c}\bar{o}} \Tstrut
\end{array}
\right] ,
\end{equation}
where
\begin{equation}\label{dec12_3}
\begin{split}
H_{h}^{12}
=&
-A_{h}^{22},
  \\
H_{h}^{22}
=&\;
A_{h}^{12}-B_{h}\mathbb{J}_{m}B_{h}^{\top}/2,
\\
H_{12}
=&\;
A_{12}-B_{h}\mathbb{J}_{m}B_{co}^{\top}\mathbb{J}_{n_{1}}/2,
\\
H_{13}
=&\;
A_{13},
\\
H_{co}
=&
-\mathbb{J}_{n_{1}}A_{co}+\mathbb{J}_{n_{1}}B_{co}\mathbb{J}_{m}B_{co}^{\top}\mathbb{J}_{n_{1}}/2,
 \\
H_{\bar{c}\bar{o}}
=&
-\mathbb{J}_{n_{2}}A_{\bar{c}\bar{o}},
\end{split}
\end{equation}
and the complex matrix $\Gamma$ must satisfy
\begin{equation}\label{Gamma}
\left[
\begin{array}{c}
\Gamma\\
\Gamma^{\#}
\end{array}
\right]
=
\left[
\begin{array}{cc|c|c}
0 & \Gamma_{h} & \Gamma_{co} & 0
\end{array}
\right] ,
\end{equation}
where
\begin{equation}\label{gamma3a}
\Gamma_{h} =V_{m}^{\dag }C_{h}, \ \ 
\Gamma_{co} =V_{m}^{\dag }C_{co}.
\end{equation}
\end{lemma}

The proof of Lemma \ref{lem:H_Gamma} is given in \textbf{Appendix}.

\begin{remark}
The matrices $\Gamma_{co}$ and $\Gamma_{h}$ in Eq. \eqref{gamma3a} are of
the form
\begin{subequations}
\begin{equation}\label{Gamma_co0}
\Gamma_{co}
=
\left[
\begin{array}{cc}
\Gamma_{co,q} & \Gamma_{co,p} \\
\Gamma_{co,q}^{\#} & \Gamma_{co,p}^{\#}
\end{array}
\right] \in \mathbb{C}^{2m\times 2n_1},
\end{equation}
and
\begin{equation}\label{Gamma_h}
\Gamma_{h} = \left[
\begin{array}{c}
\Gamma_{h,p} \\
\Gamma_{h,p}^{\#}
\end{array}
\right] \in \mathbb{C}^{2m\times n_3},
\end{equation}
\end{subequations}
respectively. Indeed, Eqs. \eqref{Gamma_co0}-\eqref{Gamma_h} can be easily established by  using Eqs. \eqref{real_Kalman_sys_A}  and (\ref{gamma3a}).
\end{remark}


\subsubsection{The sufficient condition}\label{subsec:sufficient}

We have shown in Lemma \ref{lem:H_Gamma} that if the system Hamiltonian $\boldsymbol{H}$ in Eq. (\ref{H}) and the coupling operator $\boldsymbol{L}$ in Eq. (\ref{L}) generate QSDEs in  the Kalman canonical form (\ref{real_Kalman_ss}), then the real symmetric matrix $H$ has the form (\ref{R}) and the matrix $\Gamma$ satisfies (\ref{Gamma}). In this subsection, we establish  the converse result.

\begin{lemma}\label{thm:ABC_H_Gamma}
If the real symmetric matrix $H$ for the system Hamiltonian (\ref{H}) is of the form (\ref{R}) and the complex matrix $\Gamma$ for the coupling operator (\ref{L}) satisfies Eq. (\ref{Gamma}), then the QSDEs generated are of the form (\ref{real_Kalman_ss}), with the  matrices
$\bar{A}$, $\bar{B}$ and $\bar{C}$ in Eq. (\ref{real_Kalman_sys_A}) given by
\begin{equation}\label{dec12_1}
\begin{split}
A_{h}^{11}
=&\;
H_{h}^{12^{\top}},
\\
A_{h}^{12}
=&\;
H_{h}^{22}-\imath\Gamma_{h}^{\dag }J_{m}\Gamma_{h}/2,
\\
A_{h}^{22}
=&
-H_{h}^{12},
\\
A_{12}
=&\;
H_{12}-\imath\Gamma_{h}^{\dag }J_{m}\Gamma_{co}/2,
 \\
A_{13}
=&\;
H_{13},
\\
A_{co}
=&\;
\mathbb{J}_{n_{1}}H_{co}-\imath\mathbb{J}_{n_{1}}\Gamma_{co}^{\dag}J_{m}\Gamma_{co}/2,
\\
A_{\bar{c}\bar{o}}
=&\;
\mathbb{J}_{n_{2}}H_{\bar{c}\bar{o}},
\\
A_{21}
=&\;
\mathbb{J}_{n_{1}}H_{12}^{\top}
-\imath\mathbb{J}_{n_{1}}\Gamma_{co}^{\dag }J_{m}\Gamma_{h}/2,
\\
A_{31}
=&\;
\mathbb{J}_{n_{2}}H_{13}^{\top},
\\
B_{h}
=&\;
\Gamma_{h}^{\dag }V_{m}^{\dag }\mathbb{J}_{m},
\\
B_{co}
=&\;
\mathbb{J}_{n_{1}}\Gamma_{co}^{\dag }V_{m}^{\dag }\mathbb{J}_{m},
\end{split}
\end{equation}
and
\begin{equation}  \label{dec12_2}
C_{h}
=
V_{m}\Gamma_{h}, \ \ 
C_{co}
=
V_{m}\Gamma_{co}.
\end{equation}
\end{lemma}
The proof of Lemma \ref{thm:ABC_H_Gamma}  is given in {\bf Appendix}.
%

\begin{remark}\label{rem:Hamiltonian2}
By Eq. (\ref{dec12_1}), $\mathbb{J}_{n_2}A_{\bar{c}\bar{o}} = -H_{\bar{c}\bar{o}}$ is symmetric. Thus, $A_{\bar{c}\bar{o}}$ is a Hamiltonian matrix.  Similarly, by Eq. (\ref{dec12_1}),  the matrix $\left[
\begin{array}{cc}
A_{h}^{11} & 0 \\
0 & A_{h}^{22}
\end{array}
\right]$
 is also a Hamiltonian matrix. On the other hand, if the matrix $A_{co}$ in Eq. (\ref{dec12_1}) is a Hamiltonian matrix, then
 $ \Gamma_{co}^{\dag}J_{m}\Gamma_{co}
 = \Gamma_{co}^\top J_{m}  \Gamma_{co}^{\#}$
 must hold. However, by Eq. (\ref{Gamma_co0}) it can be readily shown that
 $ \Gamma_{co}^{\dag}J_{m}\Gamma_{co}+ \Gamma_{co}^\top J_{m}  \Gamma_{co}^{\#} =0$.
 Therefore, in general the matrix $A_{co}$ is not a Hamiltonian matrix. This remark, together with Remark \ref{rem:Hamiltonian}, describes the spectral structure of the $co$, $\bar{c}\bar{o}$, and ``h'' subsystems.
\end{remark}


Given the real symmetric matrix $H$ in Eq. (\ref{R}) and complex matrix $\Gamma$ satisfying Eq. (\ref{Gamma}), Lemma \ref{thm:ABC_H_Gamma} provides a way for constructing the system matrices $\bar{A}$, $\bar{B}$ and $\bar{C}$ in the Kalman canonical form. However, to guarantee that the QSDEs are indeed the  quantum Kalman canonical form (\ref{real_Kalman_ss}), certain
controllability and observability conditions have to be satisfied.  In what follows, we investigate this problem.

We first establish the following three results, whose proofs are given in  \textbf{Appendix}.

\begin{lemma}\label{lem:equiv_1}
For the system (\ref{real_Kalman_ss}), the following statements are equivalent.

\begin{description}
\item[(i)] $( A_{h}^{11},B_{h}) $ is controllable;
\vspace{1ex}
\item[(ii)] $( A_{h}^{22},C_{h}) $ is observable;
\vspace{1ex}
\item[(iii)] $(H_{h}^{12},\Gamma_{h}) $ is observable.
\end{description}
\end{lemma}

\begin{lemma}\label{lem:equiv_2}
For the system (\ref{real_Kalman_ss}), the following statements are equivalent.
\begin{description}
\item[(i)] $( A_{co},B_{co}) $ is controllable;

\item[(ii)] $(A_{co},C_{co}) $ is observable;

\item[(iii)] $( \mathbb{J}_{n_{1}}H_{co},\Gamma_{co}) $ is observable.
\end{description}
\end{lemma}

\begin{lemma}\label{lem:equiv_mar25}
For the system (\ref{real_Kalman_ss}), the following statements are equivalent.
\begin{description}
\item[(i)]
 $\left( \left[
\begin{array}{cc}
\mathbb{J}_{n_{1}}H_{co} & 0\\
H_{12}  & H_h^{12^\top}
\end{array}
\right],  \left[\begin{array}{c}
\mathbb{J}_{n_{1}}\Gamma_{co}^\dag \\
\Gamma_h^\dag
\end{array}
\right] \right)$ is controllable;

\item[(ii)]  
 $\left( \left[
\begin{array}{cc}
\mathbb{J}_{n_{1}}H_{co} & \mathbb{J}_{n_{1}}H_{12}^\top\\
0  & -H_h^{12}
\end{array}
\right],  \left[\begin{array}{cc}
\Gamma_{co} &
\Gamma_h
\end{array}
\right] \right)$ is observable.
\end{description}
\end{lemma}

Combining Lemma \ref{lem:H_Gamma}-\ref{lem:equiv_mar25}, we
obtain the main result of this section.

\begin{theorem}\label{thm:co}
Suppose that the real matrix $H$ in Eq. (\ref{R}) and complex matrix $\Gamma$ in Eq. (\ref{Gamma}) satisfy the following conditions:

\begin{description}
\item[(i)] $H_h^{22} = H_h^{22^\top}$, $H_{\mathrm{co}} = H_{\mathrm{co}}^\top$, and
$H_{\mathrm{\bar{c}\bar{o}}} = H_{\mathrm{\bar{c}\bar{o}}}^\top$;
\vspace{1ex}
\item[(ii)] $\left( \left[
\begin{array}{cc}
\mathbb{J}_{n_{1}}H_{co} & \mathbb{J}_{n_{1}}H_{12}^\top\\
0  & -H_h^{12}
\end{array}
\right],  \left[\begin{array}{cc}
\Gamma_{co} &
\Gamma_h
\end{array}
\right] \right)$ is observable.
\end{description}
Then the resulting QSDEs are in   the Kalman canonical form (\ref{real_Kalman_ss}). In other words,
$\boldsymbol{x}_{co}$ is both controllable and observable, $\boldsymbol{x}_{\bar{c}\bar{o}}$ is neither controllable nor observable, $\boldsymbol{q}_{h}$
is controllable and unobservable, and $\boldsymbol{p}_{h} $ is
uncontrollable and observable. Conversely, if the system Hamiltonian $\boldsymbol{H}$ in Eq. (\ref{H}) and the coupling operator  $\boldsymbol{L}$ in Eq. (\ref{L}) generate  the  QSDEs in the Kalman canonical form (\ref{real_Kalman_ss}) , then the conditions (\ref{R}), (\ref{Gamma}), and (i)-(ii) above must be satisfied.
\end{theorem}

\begin{remark}\label{rem:QND}
In the Kalman canonical form  (\ref{real_Kalman_ss}), $\bold{q}_h$ is controllable and unobservable, while
$\bold{p}_h$ is observable and uncontrollable. Therefore, $\bold{p}_h$ is a vector of QND variables  (\cite{HMW95}, \cite{TC12}, \cite{NY13}, \cite{NY14}, \newline \cite{ZGPG18}). Moreover,  as shown in Lemma \ref{lem:equiv_1}, the observability of  $(A_h^{22}, C_h)$ is equivalent to the controllability of  $(A_{h}^{11},B_h)$ and both of them are equivalent to the observability of  $(H_h^{12}, \Gamma_h)$. In fact, according to
Lemma \ref{thm:ABC_H_Gamma},  the matrix pair $(H_h^{12}, \Gamma_h)$ determines the ``h'' subsystem whose quadratures $\bold{p}_h$ are QND variables. Therefore, the existence of an observable pair
$(H_h^{12}, \Gamma_h)$ generates  QND variables for  the whole quantum linear system. Interestingly, even if the pair $(H_h^{12}, \Gamma_h)$ is not an observable pair, $\bold{p}_h$ are still  QND variables provided that the condition (ii) in Theorem \ref{thm:co} holds. Examples \ref{ex:5.1_in_ZGPG18} and \ref{ex:5.2_in_ZGPG18}  illustrate this point.
\end{remark}

\subsection{A refinement of the quantum Kalman canonical form}\label{subsec:app1}

It follows from the form of the matrix $H$ in \eqref{R} that the ``h'' subsystem \eqref{sys_h1}, in general, interacts with the ``$co$'' subsystem \eqref{sys_co1_feb18} and ``$\bar{c}\bar{o}$'' subsystem \eqref{barco_system}  via the sub-matrices $H_{12}$ and $H_{13}$, respectively. As far as the Kalman canonical form is concerned, the sub-matrices $H_{12}$ and $H_{13}$ are free parameters; see also the interconnections among subsystems in Fig.~\ref{KD}. In this subsection, we  explore the structures of these and other matrices to refine the  Kalman canonical form  (\ref{real_Kalman_ss}); in particular,  we reveal its noiseless and invariant subsystems. To this end, we first introduce the following concept for quantum linear systems.

\begin{definition}\label{defn:invariant}
  If a quantum linear system $G$ can be written in a concatenation form\footnote{Given two open quantum systems $G_1\triangleq(\boldsymbol{S}_1,\boldsymbol{L}_1,\boldsymbol{H}_1)$ and $G_2\triangleq(\boldsymbol{S}_2,\boldsymbol{L}_2,\boldsymbol{H}_2)$, their concatenation product is defined to be $G_1\boxplus G_2\triangleq\left(\left[\begin{array}{cc}\boldsymbol{S}_1&0\\0&\boldsymbol{S}_2\end{array}\right],\left[\begin{array}{c}\boldsymbol{L}_1\\ \boldsymbol{L}_2\end{array}\right],\boldsymbol{H}_1+\boldsymbol{H}_2\right)$. See \cite{GJ09} for more details.}
 $G=G_1\boxplus G_2$,  then we say that  $G_1$ and $G_2$ are {\it invariant} subsystems of $G$. Moreover, an invariant subsystem is called a {\it noiseless} subsystem if it is completely isolated from the environment.
\end{definition}

Linear as well as finite-level noiseless systems and invariant systems have been studied in, e.g., \cite{TV08}, \cite{NY13}, \cite{PDP17}.

By means of the $(H,\Gamma)$ representation in Eqs. \eqref{R} and \eqref{Gamma}, we are in a position to construct the noiseless subsystem and the invariant subsystems arising in the quantum Kalman canonical form \eqref{real_Kalman_ss}. To begin with, let us consider the noiseless subsystem.

\begin{lemma}\label{lem:noiseless}
The ``$\bar{c}\bar{o}$'' subsystem \eqref{barco_system} of the Kalman canonical form \eqref{real_Kalman_ss} has a noiseless subsystem $G_{\bar{c}\bar{o}}$ if there exists an orthogonal and blockwise symplectic matrix $\mathcal{P}_{\bar{c}\bar{o}}$ such that
\begin{subequations}
\begin{equation}\label{eq:barcbaroh}
\mathcal{P}_{\bar{c}\bar{o}}\boldsymbol{x}_{\bar{c}\bar{o}}
=
 \left[\begin{array}{c}\boldsymbol{x}_{\bar{c}\bar{o}1}\\ \boldsymbol{x}_{\bar{c}\bar{o}2}\end{array}\right],  \ \mathcal{P}_{\bar{c}\bar{o}}H_{\bar{c}\bar{o}}\mathcal{P}_{\bar{c}\bar{o}}^{\top}
 =
  \left[\begin{array}{cc}H_{\bar{c}\bar{o}1}&0\\0&H_{\bar{c}\bar{o}2}\end{array}\right],
\end{equation}
\begin{equation}\label{eq:barcbaroh13}
H_{13}\mathcal{P}_{\bar{c}\bar{o}}^{\top}
=
\left[\begin{array}{cc}H_{131}&0\end{array}\right],
\end{equation}
\end{subequations}
where the system variables $\boldsymbol{x}_{\bar{c}\bar{o}1}$ and $\boldsymbol{x}_{\bar{c}\bar{o}2}$  satisfy the following CCRs
 \[
 {\rm (A1)} \ \ \left[\begin{array}{cc}\left[\begin{array}{c}\boldsymbol{x}_{\bar{c}\bar{o}1}\\ \boldsymbol{x}_{\bar{c}\bar{o}2}\end{array}\right],\left[\begin{array}{c}\boldsymbol{x}_{\bar{c}\bar{o}1}\\ \boldsymbol{x}_{\bar{c}\bar{o}2}\end{array}\right]^{\top}\end{array}\right]
=
\imath\left[\begin{array}{cc}\mathbb{J}_{n_2-n_4}&0\\0&\mathbb{J}_{n_4}\end{array}\right]
\]
 with $n_4>0$.
In this case, the noiseless subsystem $G_{\bar{c}\bar{o}}$ is given by
  \begin{equation}\label{eq:noiseless}
 \dot{\boldsymbol{x}}_{\bar{c}\bar{o}2}(t)=\mathbb{J}_{n_4}H_{\bar{c}\bar{o}2}\boldsymbol{x}_{\bar{c}\bar{o}2}(t).
\end{equation}
\end{lemma}

Due to page limitation, the proof of Lemma \ref{lem:noiseless} is omitted. However, we give the following remark.

\begin{remark}
 In order to construct a noiseless subsystem which is itself a quantum-mechanical system, the entries of $\boldsymbol{x}_{\bar{c}\bar{o}}$  need to be combined in an appropriate way, as has been done by the first equation  in  \eqref{eq:barcbaroh}. Moreover, condition (A1) in Lemma \ref{lem:noiseless} gives the CCRs for the physical quantities $\boldsymbol{x}_{\bar{c}\bar{o}1}$ and $\boldsymbol{x}_{\bar{c}\bar{o}2}$. Finally, it can be readily seen from Eqs. \eqref{eq:barcbaroh}-\eqref{eq:barcbaroh13} that  the noiseless subsystem is indeed the one in Eq. \eqref{eq:noiseless}.
\end{remark}

Compared with noiseless subsystems, general invariant subsystems are more complicated as the interaction between the quantum subsystem and the fields also needs to be considered. To make this clearer, we will study these invariant subsystems contained in the Kalman canonical form \eqref{real_Kalman_ss}.

\begin{lemma}\label{lem:coinvariant}
The ``$co$'' subsystem \eqref{sys_co1_feb18} of the Kalman canonical form \eqref{real_Kalman_ss} has an invariant subsystem $G_{co}$ if there exists an orthogonal and blockwise symplectic matrix $\mathcal{P}_{co}$ such that
\begin{equation}\label{eq:cohgamma}
\begin{split}
& \mathcal{P}_{co}\boldsymbol{x}_{co}=\left[\begin{array}{c}\boldsymbol{x}_{co1}\\ \boldsymbol{x}_{co2}\end{array}\right],\ \Gamma_{co}\mathcal{P}_{co}^{\top}=\left[\begin{array}{cc}\Gamma_{co1}&\Gamma_{co2}\end{array}\right],
\\
 &\mathcal{P}_{co}H_{co}\mathcal{P}_{co}^{\top}=\left[\begin{array}{cc}H_{co1}&0\\0&H_{co2}\end{array}\right],   H_{12}\mathcal{P}_{co}^{\top}=\left[\begin{array}{cc}H_{121}&0\end{array}\right],
\end{split}
\end{equation}
where the system variables $\boldsymbol{x}_{co1}$ and $\boldsymbol{x}_{co2}$ satisfy the condition
\begin{description}
\item[(B1)] \[\left[\begin{array}{cc}\left[\begin{array}{c}\boldsymbol{x}_{co1}\\ \boldsymbol{x}_{co2}\end{array}\right],\left[\begin{array}{c}\boldsymbol{x}_{co1}\\ \boldsymbol{x}_{co2}\end{array}\right]^{\top}\end{array}\right]=\imath\left[\begin{array}{cc}\mathbb{J}_{n_1-n_5}&0\\0&\mathbb{J}_{n_5}\end{array}\right]\] with $n_5>0$;
\end{description}
 and the constant matrices $\Gamma_{co1}$ and  $\Gamma_{co2}$ satisfy the condition
 \begin{description}
\item[(B2)] each row of $\left[\begin{array}{cc|c}\Gamma_h&\Gamma_{co1}&\Gamma_{co2}\end{array}\right]$ is in the form of either
\begin{equation*}\label{eq:Gammahco1}
\left[\begin{array}{cc|c}\Gamma_h^i&\Gamma_{co1}^i&0\end{array}\right], \ \
{\rm or} \ \
\left[\begin{array}{cc|c}0&0&\Gamma_{co2}^i\end{array}\right],
\end{equation*}
where $\Gamma_h^i$, $\Gamma_{co1}^i$ and $\Gamma_{co2}^i$, $i=1,\cdots, 2m$, are the $i$th rows of the matrices $\Gamma_h$, $\Gamma_{co1}$ and $\Gamma_{co2}$.
\end{description}
Denote the set of indices of nonzero rows of  $\Gamma_{co2}$ by  $\mathbb{I}_{co}=\{i_1,\cdots, i_{2m_1}\}$, where $m_1\leq m$, and $1\le i_1< \cdots < i_{2m_1}\le 2m$. Define
\begin{gather}
  \hat{\Gamma}_{co2}\triangleq\left[\begin{array}{c}\Gamma_{co2}^{i_1}\\ \vdots\\ \Gamma_{co2}^{i_{2m_1}}\end{array}\right],\ \boldsymbol{y}_{co}(t)\triangleq\left[\begin{array}{c}\boldsymbol{y}_{i_1}(t)\\ \vdots\\ \boldsymbol{y}_{i_{2m_1}}(t)\end{array}\right],\nonumber\\
   \boldsymbol{u}_{co}(t)\triangleq\left[\begin{array}{ccc}\boldsymbol{u}_{i_1}(t)&\cdots&\boldsymbol{u}_{i_{2m_1}}(t)\end{array}\right],\nonumber
\end{gather}
where $\boldsymbol{u}_{i}(t)$ and $\boldsymbol{y}_i(t)$ are the $i$th column and $i$th row of $\boldsymbol{u}(t)$ and  $\boldsymbol{y}(t)$, respectively. In this case, the invariant controllable and observable subsystem $G_{co}$ is given by
\begin{equation}\label{eq:coinvariant}
  \begin{split}
    \dot{\boldsymbol{x}}_{co2}(t)=&\ (\mathbb{J}_{n_5}H_{co2}-\frac{\imath}{2}\mathbb{J}_{n_5}\hat{\Gamma}_{co2}^\dag J_{m_1}\hat{\Gamma}_{co2})\boldsymbol{x}_{co2}(t)\\
    &\ +\mathbb{J}_{n_5}\hat{\Gamma}_{co2}^\dag V_{m_1}^\dag\mathbb{J}_{m_1}\boldsymbol{u}_{co}(t),\\
    \boldsymbol{y}_{co}(t)=&\ V_{m_1}\hat{\Gamma}_{co2}\boldsymbol{x}_{co2}(t)+\boldsymbol{u}_{co}(t).
  \end{split}
\end{equation}
\end{lemma}

The proof of Lemma \ref{lem:coinvariant} is given in \textbf{Appendix}.

In a similar way, we can derive the following result for the ``h'' subsystem, whose proof is omitted.

\begin{lemma}\label{cor:hinvariant}
The ``h'' subsystem \eqref{sys_h1}  of the Kalman canonical form \eqref{real_Kalman_ss}  has an invariant subsystem $G_h$ if there exists an orthogonal and blockwise symplectic matrix $\mathcal{P}_h$ such that
\begin{subequations}
\begin{equation}\label{eq:hhgamma}
\begin{split}
& \mathcal{P}_h\left[\begin{array}{c}\boldsymbol{q}_h\\ \boldsymbol{p}_h\end{array}\right]=\left[\begin{array}{c}\boldsymbol{q}_{h1}\\ \boldsymbol{p}_{h1}\\ \hline \boldsymbol{q}_{h2}\\ \boldsymbol{p}_{h2}\end{array}\right],\ \Gamma_{h}\mathcal{P}_h^{\top}=\left[\begin{array}{cc}\Gamma_{h1}&\Gamma_{h2}\end{array}\right],\\
&\mathcal{P}_h\left[\begin{array}{cc}0&H_{h}^{12}\\ H_{h}^{12^T}&H_h^{22}\end{array}\right]\mathcal{P}_h^{\top}=\left[\begin{array}{cccc}0&H_{h1}^{12}&0&0\\ H_{h1}^{12^{\top}}&H_{h1}^{22}&0&0\\0&0&0&H_{h2}^{12}\\0&0&H_{h2}^{12^{\top}}&H_{h2}^{22}\end{array}\right],
\end{split}
\end{equation}
\begin{equation}\label{eq:h12h13}
\mathcal{P}_h\left[\begin{array}{c}0\\H_{12}\end{array}\right]=\left[\begin{array}{c}0\\H_{12}^1\\0\\0\end{array}\right],\ \mathcal{P}_h\left[\begin{array}{c}0\\H_{13}\end{array}\right]=\left[\begin{array}{c}0\\H_{13}^1\\0\\0\end{array}\right],
\end{equation}
\end{subequations}
where the system variables  $\boldsymbol{q}_{h1}$, $\boldsymbol{p}_{h1}$, $\boldsymbol{q}_{h2}$,  and $\boldsymbol{p}_{h2}$ satisfy the conditions
\begin{description}
  \item[(C1)] \[\left[\left[\begin{array}{c}\boldsymbol{q}_{h1}\\ \boldsymbol{p}_{h1}\\ \boldsymbol{q}_{h2}\\ \boldsymbol{p}_{h2}\end{array}\right],\left[\begin{array}{c}\boldsymbol{q}_{h1}\\ \boldsymbol{p}_{h1}\\ \boldsymbol{q}_{h2}\\ \boldsymbol{p}_{h2}\end{array}\right]^{\top}\right]=\imath\left[\begin{array}{cc}\mathbb{J}_{n_3-n_6}&0\\0&\mathbb{J}_{n_6}\end{array}\right]\] with $n_6>0$;
 \end{description}
  and constant matrices $\Gamma_{h1}$ and $\Gamma_{h2}$ satisfy the condition
   \begin{description}
  \item[(C2)] each row of $\left[\begin{array}{cc|c}\Gamma_{h1}&\Gamma_{co}&\Gamma_{h2}\end{array}\right]$ is in the form
\begin{equation*}\label{eq:Gammah1co}
\left[\begin{array}{cc|c}\Gamma_{h1}^i&\Gamma_{co}^i&0\end{array}\right], \ \
{\rm or} \ \
\left[\begin{array}{cc|c}0&0&\Gamma_{h2}^i\end{array}\right],
\end{equation*}
where $\Gamma_{h1}^i$, $\Gamma_{h2}^i$ and $\Gamma_{co}^i$, $i=1,\cdots, 2m$, are the $i$th rows of the matrices $\Gamma_{h1}$, $\Gamma_{h2}$ and $\Gamma_{co}$.
\end{description}
Denote the set of indices of nonzero rows of $\Gamma_{h2}$ by  $\mathbb{I}_{h}=\{i_1,\cdots, i_{2m_2}\}$,  where $m_2\leq m$ and $1\le i_1 < \cdots < i_{2m_2}\leq 2m$. Define
\begin{gather}
  \hat{\Gamma}_{h2}\triangleq\left[\begin{array}{c}\Gamma_{h2}^{i_1}\\ \vdots\\ \Gamma_{h2}^{i_{2m_2}}\end{array}\right],\ \boldsymbol{y}_{h}(t)\triangleq\left[\begin{array}{c}\boldsymbol{y}_{i_1}(t)\\ \vdots\\ \boldsymbol{y}_{i_{2m_2}}(t)\end{array}\right],\nonumber\\
   \boldsymbol{u}_{h}(t)\triangleq\left[\begin{array}{ccc}\boldsymbol{u}_{i_1}(t)&\cdots&\boldsymbol{u}_{i_{2m_2}}(t)\end{array}\right],\nonumber
\end{gather}
where $\boldsymbol{u}_{i}(t)$ and $\boldsymbol{y}_i(t)$ are the $i$th column  and $i$th row of $\boldsymbol{u}(t)$ and $\boldsymbol{y}(t)$, respectively. Then the invariant subsystem $G_{h}$ is given by
\begin{equation}\label{eq:hinvariant}
  \begin{split}
    \left[\begin{array}{c}\dot{\boldsymbol{q}}_{h2}(t)\\ \dot{\boldsymbol{p}}_{h2}(t)\end{array}\right]
    =&
    \ \left[\begin{array}{cc}H_{h2}^{12^\top}&H_{h2}^{22}-\frac{\imath}{2}\hat{\Gamma}_{h2}^\dag J_{m_2}\hat{\Gamma}_{h2}
    \\
    0&-H_{h2}^{12} \Tstrut
    \end{array}\right] \left[\begin{array}{c}\boldsymbol{q}_{h2}(t)\\ \boldsymbol{p}_{h2}(t)\end{array}\right]\\
    &\ +\left[\begin{array}{c}\hat{\Gamma}_{h2}^\dag V_{m_2}^\dag\mathbb{J}_{m_2}\\0\end{array}\right]\boldsymbol{u}_{h}(t),\\
    \boldsymbol{y}_{h}(t)=&\ V_{m_2}\hat{\Gamma}_{h2}\boldsymbol{p}_{h2}(t)+\boldsymbol{u}_{h}(t).
  \end{split}
\end{equation}
\end{lemma}

By removing subsystems $G_{\bar{c}\bar{o}}, G_{co},G_h$ from the Kalman canonical form \eqref{real_Kalman_ss}, the remaining subsystem, denoted by $G_m$, is clearly an invariant subsystem. Next, we  introduce this invariant subsystem. Let $m_3=m-m_1-m_2$.  Denote  the set of indices of nonzero rows of $[\begin{array}{cc}\Gamma_{h1}&\Gamma_{co1}\end{array}]$ by $\mathbb{I}_{m}=\{i_1,\cdots, i_{2m_3}\}$, $1\le i_1< \cdots < i_{2m_3}\le 2m$. Define
\begin{gather}
  \left[\hat{\Gamma}_{h1}\ \hat{\Gamma}_{co1}\right]\triangleq\left[\begin{array}{cc}\Gamma_{h1}^{i_1}&\Gamma_{co1}^{i_1}
  \\
  \vdots & \vdots\\
   \Gamma_{h1}^{i_{2m_3}}&\Gamma_{co1}^{i_{2m_3}}\end{array}\right],\  
   \boldsymbol{x}_{m}(t)\triangleq \left[\begin{array}{c}\boldsymbol{q}_{h1}(t)\\ \boldsymbol{p}_{h1}(t)\\ \hline\boldsymbol{x}_{co1}(t)\\ \hline\boldsymbol{x}_{\bar{c}\bar{o}1}(t)\end{array}\right]
     \nonumber
  \\
    \boldsymbol{y}_{m}(t)\triangleq\left[\begin{array}{c}\boldsymbol{y}_{i_1}(t)\\
    \vdots\\ \boldsymbol{y}_{i_{2m_3}}(t)\end{array}\right], \ 
   \boldsymbol{u}_{m}(t)\triangleq\left[\boldsymbol{u}_{i_1}(t)\
    \cdots\ \boldsymbol{u}_{i_{2m_3}}(t)\right],
   \nonumber
\end{gather}
where $\Gamma_{h1}^i$, $\Gamma_{co1}^i$, $i=1,\cdots, 2m$, are the $i$th row of matrices $\Gamma_{h1}$, $\Gamma_{co1}$, and $\boldsymbol{u}_{i}(t)$, $\boldsymbol{y}_i(t)$ are the $i$th column  and $i$th row of $\boldsymbol{u}(t)$ and $\boldsymbol{y}(t)$, respectively. The invariant subsystem $G_{m}$ is of the form
\begin{equation}\label{eq:minvariant}
  \begin{split}
\dot{\boldsymbol{x}}_{m}(t)
    =&\ 
    \vec{A}\boldsymbol{x}_{m}(t) +\vec{B}\boldsymbol{u}_{m}(t),\\
    \boldsymbol{y}_{m}(t)=&\ \vec{C}\boldsymbol{x}_{m}(t) +\boldsymbol{u}_{m}(t),
  \end{split}
\end{equation}
where
\begin{gather}
  \vec{A}\triangleq\left[\begin{array}{cc|c|c}A_{h1}^{11}&A_{h1}^{12}&A_{m12}&A_{m13}\\
    0&A_{h1}^{22}&0& 0 \Tstrut \\
    \hline0&A_{m21}&A_{co1}&0 \Tstrut \\ \hline0&A_{m31}&0&A_{\bar{c}\bar{o}1} \Tstrut\end{array}\right],\ \vec{B}\triangleq\left[\begin{array}{c}B_{h1}\\0\\ \hline B_{co1}\\ \hline0\end{array}\right],\nonumber\\
    \vec{C}\triangleq\left[\begin{array}{cc|c|c}0&C_{h1}&C_{co1}&0\end{array}\right],\nonumber
\end{gather}
with $A_{h1}^{11}=H_{h1}^{12^{\top}}$,
     $A_{h1}^{12}= H_{h1}^{22}-\frac{\imath}{2}\hat{\Gamma}_{h1}^\dag J_{m_3}\hat{\Gamma}_{h1}$,
    $A_{h1}^{22}=-H_{h1}^{12}$,
     $ A_{m12}= H_{m12}-\frac{\imath}{2}\hat{\Gamma}_{h1}^\dag J_{m_3}\hat{\Gamma}_{co1}$,
     $A_{m13}= H_{m13}$,
    $A_{co1}= \mathbb{J}_{n_1-n_5}H_{co1}-\frac{\imath}{2}\mathbb{J}_{n_1-n_5}\hat{\Gamma}_{co1}^\dag J_{m_3}\hat{\Gamma}_{co1}$,
    $A_{\bar{c}\bar{o}1}= \mathbb{J}_{n_2-n_4}H_{\bar{c}\bar{o}1}$,
    $A_{m21}= \mathbb{J}_{n_1-n_5}H_{m12}^{\top}-\frac{\imath}{2}\mathbb{J}_{n_1-n_5}\hat{\Gamma}_{co1}^\dag J_{m_3}\hat{\Gamma}_{h1}$,
    $A_{m31}= \mathbb{J}_{n_2-n_4}H_{m13}^{\top}$,
    $B_{h1}=\hat{\Gamma}_{h1}^\dag V_{m_3}\mathbb{J}_{m_3}$,
    $B_{co1}=\mathbb{J}_{n_1-n_5}\hat{\Gamma}_{co1}^\dag V_{m_3}\mathbb{J}_{m_3}$,
    $C_{h1}= V_{m_3}\hat{\Gamma}_{h1}$,
    and $C_{co1}=V_{m_3}\hat{\Gamma}_{co1}$,

Based on Lemmas \ref{lem:noiseless}-\ref{cor:hinvariant} and the subsystem $G_m$ given in \eqref{eq:minvariant}, we are now in a position to propose the following result.

\begin{figure}
  \centering
  \includegraphics[width=0.6\textwidth]{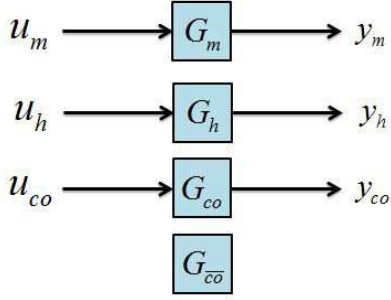}\\
  \caption{Block diagram for the quantum Kalman canonical form  $G= G_{\bar{c}\bar{o}} \boxplus G_{co} \boxplus G_h \boxplus   G_m $, as given in Theorem \ref{thm:generalresult}.} \label{completesystem}
\end{figure}

\begin{theorem}\label{thm:generalresult}
The quantum Kalman canonical form \eqref{real_Kalman_ss} can be put in the concatenation form
  \begin{equation}\label{eq:complete}
    G= G_{\bar{c}\bar{o}} \boxplus G_{co} \boxplus G_h \boxplus   G_m,
  \end{equation}
  where $G_{\bar{c}\bar{o}}$ is the noiseless subsystem given in Lemma \ref{lem:noiseless}, $G_{co}$  is the invariant controllable and observable subsystem given in  Lemma \ref{lem:coinvariant},  $G_h$ is the invariant subsystem given in  Lemma \ref{cor:hinvariant}, and $G_m$ is given in \eqref{eq:minvariant},  provided that
 \begin{description}
 \item[(i)] there exist orthogonal and blockwise symplectic matrices $\mathcal{P}_{\bar{c}\bar{o}}$, $\mathcal{P}_{co}$ and $\mathcal{P}_{h}$ satisfying \eqref{eq:barcbaroh}, \eqref{eq:cohgamma}, \eqref{eq:hhgamma}, and
 \begin{equation*}\label{eq:h12h13final}
  \begin{split}
    \mathcal{P}_h\left[\begin{array}{c}0\\H_{12}\end{array}\right]\mathcal{P}_{co}^{\top}=&\ \left[\begin{array}{cc}0&0\\H_{m12}&0\\0&0\\0&0\end{array}\right],\\ \mathcal{P}_h\left[\begin{array}{c}0\\H_{13}\end{array}\right]\mathcal{P}_{\bar{c}\bar{o}}^{\top}=&\ \left[\begin{array}{cc}0&0\\H_{m13}&0\\0&0\\0&0\end{array}\right];
  \end{split}
\end{equation*}
  \item[(ii)] (A1), (B1)-(B2), and (C1)-(C2) hold;
  \item[(iii)]  each row of $\left[\begin{array}{cc|c|c}\Gamma_{h1}&\Gamma_{co1}&\Gamma_{h2}&\Gamma_{co2}\end{array}\right]$ is in one of the following forms
\begin{subequations}
\begin{equation*}\label{vecgamma1_feb11}
\left[\begin{array}{cc|c|c}\Gamma^i_{h1}&\Gamma^i_{co1}&0&0\end{array}\right],
\end{equation*}
or
\begin{equation*}\label{vecgamma2_feb11}
\left[\begin{array}{cc|c|c}0&0&\Gamma^i_{h2}&0\end{array}\right],
\end{equation*}
or
\begin{equation*}\label{vecgamma2b_feb11}
\left[\begin{array}{cc|c|c}0&0&0&\Gamma^i_{co2}\end{array}\right],
\end{equation*}
\end{subequations}
where $\Gamma_{h1}^i$, $\Gamma_{h2}^i$, $\Gamma_{co1}^i$ and $\Gamma_{co2}^i$, $i=1,\cdots, 2m$, are the $i$th row of matrices $\Gamma_{h1}$, $\Gamma_{h2}$, $\Gamma_{co1}$ and $\Gamma_{co2}$.
\end{description}

A block diagram for a quantum linear system in the form \eqref{eq:complete} is shown in Fig. \ref{completesystem}.
\end{theorem}

\begin{remark}\label{rem:main}
  We have the following observations on Theorem \ref{thm:generalresult}.
\begin{description}
\item[(i)] The noiseless subsystem $G_{\bar{c}\bar{o}}$ is a subsystem of the $\bar{c}\bar{o}$ subsystem, as can be seen from Eq. (\ref{eq:noiseless}); similarly, the invariant subsystem  $G_{co} $ is a subsystem of the $co$ subsystem, as can be seen from Eq. (\ref{eq:coinvariant}); and the invariant subsystem  $G_h$ is a subsystem of the ``h'' subsystem,  as can be seen from Eq. (\ref{eq:hinvariant});

\item[(ii)] The invariant subsystem  $G_m$ is a mixture of ``h'', $co$, and $\bar{c}\bar{o}$ subsystems, as can be seen from Eq. (\ref{eq:minvariant}).

\item[(iii)] Comparison of Figs. \ref{KD} and \ref{completesystem} tells us the  system \eqref{eq:complete} in Theorem \ref{thm:generalresult} involves partitioning  system inputs and outputs, while the original Kalman canonical form \eqref{real_Kalman_ss} does not.

\item[(iv)] It is worthwhile to notice that the system decomposition in Fig.~\ref{completesystem} is very general.  In some cases,  some of the subsystems in Fig.~\ref{completesystem} may not exist; this can be easily seen from the conditions in Lemmas \ref{lem:noiseless}-\ref{cor:hinvariant} and Theorem \ref{thm:generalresult}.  On the other hand, there might be more than one invariant $co$, $\bar{c}\bar{o}$, or  ``h'' subsystems. Indeed, the system in  Example \ref{ex:Thm5.1_2} below can be decomposed into two invariant $co$ subsystems, each of which is a harmonic oscillator driven by a single input field.
\end{description}
 It is interesting to see that the subsystem $G_m$ is in the Kalman canonical form \eqref{real_Kalman_ss}, while $G_{\bar{c}\bar{o}}$, $G_h$, and $G_{co} $ are in the form of \eqref{barco_system}, \eqref{sys_h1}, and \eqref{sys_co1_feb18} respectively.  Therefore, the quantum Kalman canonical form \eqref{real_Kalman_ss} is decomposed into four subsystems which are decoupled from each other, and one of which itself is a smaller Kalman canonical form. This means that the Kalman canonical form \eqref{real_Kalman_ss}, in general, may not reveal the noiseless subsystem and the invariant subsystems of a given quantum linear system. In Theorem \ref{thm:generalresult}, a refined decomposition of the matrices $H$ and $\Gamma$  shows that a  quantum linear system can be expressed in the form \eqref{eq:complete} by appropriate coordinate transformations. Finally, the following should be noted. In  the Heisenberg picture of quantum mechanics, a quantum linear system $G$ may be put into a concatenation form, as shown in Fig.~\ref{completesystem}, where four possible subsystems are decoupled form each other.  However, the initial state of the whole system $G$ may still be a state superposed among  all these  subsystems.
\end{remark}

\section{Applications to quantum BAE measurements}\label{sec:application}

In this section, we consider the realization of  BAE measurements. We present necessary and sufficient conditions for the quantum Kalman canonical form (\ref{real_Kalman_ss}) to realize BAE measurements. These necessary and sufficient conditions are given explicitly in terms of the  physical parameters $H$ and $\Gamma$.

It is mentioned in  Remark \ref{rem:QND}  that QND variables are related to the  ``h'' subsystem in the Kalman canonical form (\ref{real_Kalman_ss}). In contrast, as BAE measurements are an input-output property, they are determined completely by the $co$ subsystem. For the quantum linear system (\ref{real_Kalman_ss}),
the transfer function from $\boldsymbol{u}$ to $\boldsymbol{y}$ is
\begin{equation*}
\Xi _{\boldsymbol{u}\rightarrow \boldsymbol{y}}(s)
=
C_{co}(sI-A_{co})^{-1}B_{co}+I.
\end{equation*}
Partition the matrices $B_{co}$ and $C_{co}$ as
\[
B_{co}= [B_{co,q} \  B_{co,p}], \ \
C_{co}=\left[
\begin{array}{c}
C_{co,q} \\
C_{co,p}
\end{array}
\right],
\]
respectively. Then the transfer function from $\boldsymbol{p}_{\mathrm{in}}$
to $\boldsymbol{q}_{\mathrm{out}}$ is
\begin{equation}\label{tf:pq}
\Xi _{\boldsymbol{p}_{\mathrm{in}}\rightarrow \boldsymbol{q}_{\mathrm{out}}}(s)
=
C_{co,q}(sI-A_{co})^{-1}B_{co,p}.
\end{equation}
Similarly, the transfer function from $\boldsymbol{q}_{\mathrm{in}}$ to $
\boldsymbol{p}_{\mathrm{out}}$ is
\begin{equation*}\label{tf:qp}
\Xi _{\boldsymbol{q}_{\mathrm{in}}\rightarrow \boldsymbol{p}_{\mathrm{out}}}(s)
=
C_{co,p}(sI-A_{co})^{-1}B_{co,q}.
\end{equation*}

The following is the main result of this section, which gives necessary and sufficient
conditions for the realization of  BAE measurements by the quantum linear system (\ref{real_Kalman_ss}).

\begin{theorem} \label{thm:BAE}
\begin{description}
\item[(i)] The quantum Kalman canonical form (\ref{real_Kalman_ss}) realizes the BAE measurements of $\boldsymbol{q}_{\mathrm{out}}$ with
respect to $\boldsymbol{p}_{\mathrm{in}}$; i.e.,
\begin{equation} \label{tf:pq0}
\Xi _{\boldsymbol{p}_{\mathrm{in}}\rightarrow \boldsymbol{q}_{\mathrm{out}}}(s)\equiv 0 
\end{equation}
if and only if
\begin{eqnarray}
&&\left[
\begin{array}{cc}
\mathrm{Re}\left( \Gamma_{co,q}\right) & \mathrm{Re}\left( \Gamma_{co,p}\right)
\end{array}
\right] (sI-\mathbb{J}_{n_{1}}H_{co})^{-1}
\nonumber
\\
&& \times \left[
\begin{array}{c}
\mathrm{Re}(\Gamma_{co,p}^{\top})
\vspace{3pt}
 \\
-\mathrm{Re}(\Gamma_{co,q}^{\top}) %
\end{array}
\right]\equiv0;  \label{temp9}
\end{eqnarray}
\item[(ii)] The quantum Kalman canonical form (\ref{real_Kalman_ss}) realizes the BAE measurements of $\boldsymbol{p}_{\mathrm{out}}$ with
respect to $\boldsymbol{q}_{\mathrm{in}}$; i.e.,
\begin{equation*}\label{tf:pq1}
\Xi _{\boldsymbol{q}_{\mathrm{in}}\rightarrow \boldsymbol{p}_{\mathrm{out}
}}(s)\equiv0  
\end{equation*}
if and only if
\begin{eqnarray}
&&\left[
\begin{array}{cc}
\mathrm{Im}\left( \Gamma_{co,q}\right) & \mathrm{Im}\left( \Gamma_{co,p}\right)
\end{array}
\right] (sI-\mathbb{J}_{n_{1}}H_{co})^{-1}
\nonumber
\\
&& \times \left[
\begin{array}{c}
\mathrm{Im}(\Gamma_{co,p}^{\top})
\vspace{3pt}
\\
-\mathrm{Im}( \Gamma_{co,q}^{\top})%
\end{array}
\right]\equiv0.
  \label{temp9b}
\end{eqnarray}
\end{description}
\end{theorem}

The proof of Theorem \ref{thm:BAE} is given in \textbf{Appendix}.

The following corollary presents a special case of Theorem \ref{thm:BAE}.

\begin{corollary}\label{cor:BAE}
Let a quantum linear system be parametrized by  the Hamiltonian
$\boldsymbol{H}=\frac{1}{2}\boldsymbol{x}^{\top}H\boldsymbol{x}$  and  the
coupling operator
$\boldsymbol{L}=\Gamma\boldsymbol{x}$,
where  $\boldsymbol{x}$ satisfies the CCRs $[\boldsymbol{x}, \ \boldsymbol{x}^\top]=\imath \mathbb{J}_n$,
\begin{equation}
H=\left[
\begin{array}{cc}
0 & I_{n} \\
I_n & 0
\end{array}
\right], \label{H_co}
\end{equation}
and
\begin{equation*}
\Gamma = [\Gamma_q \  \ \Gamma_p]
\end{equation*}
 with $\Gamma_q, \Gamma_p\in \mathbb{C}^{m\times n}$. We have:

\begin{description}
\item[(i)] The system is controllable and observable  and $\Xi _{\boldsymbol{p}_{\mathrm{in}}\rightarrow \boldsymbol{q}_{\mathrm{out}}}(s)\equiv0$ if and only if
\begin{eqnarray}
\mathrm{Re}\left( \Gamma_{q}\right) \perp \mathrm{Re}\left( \Gamma_{p}\right) ,
\label{perp_1a}
\end{eqnarray}
and
\begin{equation}
\mathrm{rank}\left( \left[
\begin{array}{c}
\breve{\Gamma} \\
\breve{\Gamma}J_{n }
\end{array}
\right] \right) =2n ,  \label{rank Gamma}
\end{equation}
where $\breve{\Gamma} \triangleq \left[\begin{array}{c}\Gamma \\ \Gamma^\# \end{array}\right]$;
\item[(ii)] The system is controllable and observable  and
$\Xi _{\boldsymbol{q}_{\mathrm{in}}\rightarrow \boldsymbol{p}_{\mathrm{out}}}(s)\equiv 0$ if and only if
\begin{equation}
\mathrm{Im}\left( \Gamma_{q}\right) \perp \mathrm{Im}\left( \Gamma_{p}\right) ,
\label{perp_2a}
\end{equation}
and (\ref{rank Gamma}) hold.
\end{description}
\end{corollary}

The proof of Corollary \ref{cor:BAE} is given in  \textbf{Appendix}.

\section{Examples}\label{sec:example}
In this section, we use several example to illustrate the theoretical results derived in this paper.

The first example is used to illustrate Theorems \ref{thm:co} and \ref{thm:generalresult}.
\begin{example}\label{ex:5.1_in_ZGPG18} This example is taken from Example 5.1 of \cite{ZGPG18}, where three scenarios, red-detuned regime, blue-detuned regime and phase-shift regimes, have been investigated for a quantum opto-mechanical system. It is easy to see that in the red-detuned regime, the quantum system can be decomposes as  $G_{co} \boxplus G_{\bar{c}\bar{o}}$, the same is true for the blue-detuned regime. However, in the phase-shift regime,  the quantum system can be decomposes as  $G_m \boxplus G_{\bar{c}\bar{o}}$. According to \cite[Eq. (83)]{ZGPG18} and Lemma \ref{lem:H_Gamma} in Subsection \ref{subsec:necessary}, we can get the following parameters: 
$H_h^{12} = 0$, $H_h^{22} = 0$, $H_{12}= \lambda[1 \ \ 0]$, $H_{13}=0$, $H_{co}=\omega_m\begin{smallmatrix}\left[\begin{array}{cc}
1 &0\\
0 &-1
\end{array}
\right]\end{smallmatrix}$, $H_{\bar{c}\bar{o}}=0$, $\Gamma_h=0$, $\Gamma_{co}  =\sqrt{\frac{\kappa}{2}}  \begin{smallmatrix}\left[\begin{array}{cc}
1 &\imath\\
1 &-\imath
\end{array}
\right]\end{smallmatrix}$. By Eq. (\ref{Gamma_co0}), $\Gamma_{co,q}=\sqrt{\frac{\kappa}{2}} $ and  $\Gamma_{co,p}=\sqrt{\frac{\kappa}{2}} \imath$. It can be easily verified that Theorem \ref{thm:co} holds.
\end{example}

The following example is used to illustrate Theorems \ref{thm:co}  and \ref{thm:BAE}.
\begin{example}\label{ex:5.2_in_ZGPG18} This example is taken from Example 5.2 of \cite{ZGPG18}, which discussed a real physical experiment considered in \cite{ODP+16} and \cite{WC13}. According to \cite[Eq. (83)]{ZGPG18} and Lemma \ref{lem:H_Gamma}, we can get the following parameters: 
$H_h^{12} = -\Omega \mathbb{J}_1$, $H_h^{22} = 0$, $H_{12}= 2\sqrt{2}g\begin{smallmatrix}\left[\begin{array}{cc}
0 &0\\
1 &0
\end{array}
\right]\end{smallmatrix}$, $H_{13}=0$, $H_{co}=0$, $H_{\bar{c}\bar{o}}=0$, $\Gamma_h=0$, $\Gamma_{co} =\sqrt{\frac{\kappa}{2}} \begin{smallmatrix}\left[\begin{array}{cc}
1 &\imath\\
1 &-\imath
\end{array}
\right]\end{smallmatrix}$. By Eq. (\ref{Gamma_co0}), $\Gamma_{co,q}=\sqrt{\frac{\kappa}{2}} $ and  $\Gamma_{co,p}=\sqrt{\frac{\kappa}{2}} \imath$. t can be easily verified that Theorem \ref{thm:co} holds.
Moreover, it is straightforward to very that both Eqs. (\ref{temp9}) and  (\ref{temp9b}) in Theorem \ref{thm:BAE} hold.  Indeed, as analyzed in \cite[Example 5.2]{ZGPG18}, this system realizes a quantum BAE measurement of with
respect to $\boldsymbol{p}_{\mathrm{in}}$ and  also  that of  $\boldsymbol{p}_{\mathrm{out}}$ with respect to $\boldsymbol{q}_{\mathrm{in}}$.
\end{example}

The third example is used to illustrate Corollary \ref{cor:BAE}.
\begin{example}
Let $n =m=1$. Choose
\begin{equation*}
H=\left[
\begin{array}{cc}
0 & 1 \\
1 & 0
\end{array}
\right] , \ \ \Gamma_{q}=\imath, \ \ \Gamma_{p}=-\imath.
\end{equation*}
Clearly, $H$ is of the form (\ref{H_co}), and Eq. (\ref{perp_1a}) is satisfied. Moreover, Eq. (\ref{rank Gamma}) holds, but   Eq. (\ref{perp_2a}) does not.
In fact, with the above system parameters,  it is easy to see  that this controllable and observable system is described by the QSDEs:
\begin{align*}
\left[
\begin{array}{c}
\boldsymbol{\dot{q}}(t)\\
\boldsymbol{\dot{p}}(t)
\end{array}
\right] =&\;\left[
\begin{array}{cc}
1 & 0 \\
0 & -1
\end{array}
\right]\left[
\begin{array}{c}
\boldsymbol{q}(t)\\
\boldsymbol{p}(t)
\end{array}
\right]
+\sqrt{2}\left[
\begin{array}{cc}
1 & 0 \\
1 & 0
\end{array}
\right] \left[
\begin{array}{c}
\boldsymbol{q}_{\mathrm{in}}(t) \\
\boldsymbol{p}_{\mathrm{in}}(t)
\end{array}
\right],
\\
\left[
\begin{array}{c}
\boldsymbol{q}_{\mathrm{out}}(t) \\
\boldsymbol{p}_{\mathrm{out}}(t)
\end{array}
\right] =&\;\sqrt{2}\left[
\begin{array}{cc}
0 & 0 \\
1 & -1%
\end{array}
\right] \left[
\begin{array}{c}
\boldsymbol{q}(t)\\
\boldsymbol{p}(t)
\end{array}
\right]+\left[
\begin{array}{c}
\boldsymbol{q}_{\mathrm{in}}(t) \\
\boldsymbol{p}_{\mathrm{in}}(t)
\end{array}
\right] .
\end{align*}
It can be verified that
\[
\Xi _{\boldsymbol{p}_{\mathrm{in}}\rightarrow \boldsymbol{q}_{\mathrm{out}
}}(s) \equiv 0, \ \
\Xi _{\boldsymbol{q}_{\mathrm{in}}\rightarrow \boldsymbol{p}_{\mathrm{out}
}}(s)  = \frac{2}{s-1}-\frac{2}{s+1} \neq 0.
\]
Finally, for this system, the Hamiltonian $\boldsymbol{H}$ and the coupling operator $\boldsymbol{L}$ are respectively
\begin{align*}
\boldsymbol{H}
=&
\;\frac{\boldsymbol{q}\boldsymbol{p}
+\boldsymbol{p}\boldsymbol{q}}{2}
=\frac{\boldsymbol{a}_{co}^{2}-(\boldsymbol{a}_{co}^{\ast })^{2}}{2\imath},
\\
\boldsymbol{L}
=&
\;\imath(\boldsymbol{p}-\boldsymbol{q})
=\frac{1+\imath}{\imath\sqrt{2}}\boldsymbol{a}_{co}
-\left( \frac{1+\imath}{\imath\sqrt{2}}\right) ^{\ast }\boldsymbol{a}_{co}^{\ast }.
\end{align*}
This system can be physically realized by means of quantum optical devices; see, e.g., \cite{JNP08}, \cite{NJD09}, \cite{HIN14}.
\end{example}

The final example is used to illustrate Theorems \ref{thm:generalresult} and \ref{thm:BAE} and  Remark \ref{rem:main}.
\begin{example}\label{ex:Thm5.1_2}
This example considers the Michelson's interferometer which is one of the simplest devices for  gravitational wave detection, see \cite[Fig.\;3(c)]{NY14}. The interferometer contains two identical mechanical oscillators with position quadratures $\boldsymbol{q}_1$, $\boldsymbol{q}_2$, and momentum quadratures $\boldsymbol{p}_1$, $\boldsymbol{p}_2$, respectively.  The resonant frequency and mass of the mechanical oscillators are denoted by $\omega_m$ and $m$, respectively. The input coherent light field (the probe field $\hat{W}_1$ in \cite[Fig.\;3(c)]{NY14}) and the input vacuum light field ($\hat{W}_2$ in \cite[Fig.\;3(c)]{NY14}) are described by their respective position quadratures $\boldsymbol{q}_{\mathrm{in},1}$, $\boldsymbol{q}_{\mathrm{in},2}$, and momentum quadratures $\boldsymbol{p}_{\mathrm{in},1}$, $\boldsymbol{p}_{\mathrm{in},2}$.  Let $\lambda$ be the coupling strength between the probe field and the mechanical oscillators. It is assumed that the mechanical oscillators are subjected to forces $F$ and $-F$. Then the dynamics of the system, given in \cite[Eq. (19)]{NY14},  is described by the following QSDEs.
  \begin{equation}\label{eq:interferometer}
    \begin{split}
      \left[\begin{array}{c}\boldsymbol{\dot{q}}_1\\ \boldsymbol{\dot{q}}_2\\ \boldsymbol{\dot{p}}_1\\ \boldsymbol{\dot{p}}_2\end{array}\right]=&\ \left[\begin{array}{cccc}0&0&1/m&0\\0&0&0&1/m\\-m\omega_m^2&0&0&0\\0&-m\omega_m^2&0&0\end{array}\right]\left[\begin{array}{c}\boldsymbol{q}_1\\ \boldsymbol{q}_2\\ \boldsymbol{p}_1\\ \boldsymbol{p}_2\end{array}\right]\\
      &\ +\sqrt{\lambda}\left[\begin{array}{cccc}0&0&0&0\\0&0&0&0\\1&1&0&0\\1&-1&0&0\end{array}\right]\left[\begin{array}{c}\boldsymbol{q}_{\mathrm{in},1}\\ \boldsymbol{q}_{\mathrm{in},2}\\ \boldsymbol{p}_{\mathrm{in},1}\\ \boldsymbol{p}_{\mathrm{in},2}\end{array}\right],\\
      \left[\begin{array}{c}\boldsymbol{q}_{\mathrm{out},1}\\ \boldsymbol{q}_{\mathrm{out},2}\\ \boldsymbol{p}_{\mathrm{out},1}\\ \boldsymbol{p}_{\mathrm{out},2}\end{array}\right]=&\ \sqrt{\lambda}\left[\begin{array}{cccc}0&0&0&0\\0&0&0&0\\1&1&0&0\\1&-1&0&0\end{array}\right]\left[\begin{array}{c}\boldsymbol{q}_1\\ \boldsymbol{q}_2\\ \boldsymbol{p}_1\\ \boldsymbol{p}_2\end{array}\right]+\left[\begin{array}{c}\boldsymbol{q}_{\mathrm{in},1}\\ \boldsymbol{q}_{\mathrm{in},2}\\ \boldsymbol{p}_{\mathrm{in},1}\\ \boldsymbol{p}_{\mathrm{in},2}\end{array}\right].
    \end{split}
  \end{equation}
    Note that the system \eqref{eq:interferometer} is both controllable and observable. It follows from Lemma \ref{lem:H_Gamma} that
\begin{equation}\label{interferometer:hgamma}
\begin{split}
  H=&\ \left[\begin{array}{cccc}m\omega_m^2&0&0&0\\0&m\omega_m^2&0&0\\0&0&1/m&0\\0&0&0&1/m\end{array}\right],\\
  \Gamma=&\ \sqrt{\frac{\lambda}{2}}\left[\begin{array}{cccc}\imath&\imath&0&0\\ \imath&-\imath&0&0\end{array}\right].
\end{split}
\end{equation}
It is easy to check that $H$ and $\Gamma$ in Eq. \eqref{interferometer:hgamma} satisfy the condition \eqref{temp9}, but not the condition \eqref{temp9b}. Indeed, denote $\boldsymbol{q}_{\mathrm{in}}=\left[\begin{array}{c}\boldsymbol{q}_{\mathrm{in},1}\\ \boldsymbol{q}_{\mathrm{in},2}\end{array}\right]$,  $\boldsymbol{p}_{\mathrm{in}}=\left[\begin{array}{c}\boldsymbol{p}_{\mathrm{in},1}\\ \boldsymbol{p}_{\mathrm{in},2}\end{array}\right]$, $\boldsymbol{q}_{\mathrm{out}}=\left[\begin{array}{c}\boldsymbol{q}_{\mathrm{out},1}\\ \boldsymbol{q}_{\mathrm{out},2}\end{array}\right]$, and $\boldsymbol{p}_{\mathrm{out}}=\left[\begin{array}{c}\boldsymbol{p}_{\mathrm{out},1}\\ \boldsymbol{p}_{\mathrm{out},2}\end{array}\right]$. It turns out that
\begin{align*}
\Xi _{\boldsymbol{p}_{\mathrm{in}}\rightarrow \boldsymbol{q}_{\mathrm{out}}}(s) \equiv 0 , \ \
\Xi _{\boldsymbol{q}_{\mathrm{in}}\rightarrow \boldsymbol{p}_{\mathrm{out}}}(s)=\frac{2}{m(s^2+\omega_m^2)} I_2  \neq 0 .
\end{align*}
Therefore the QSDEs \eqref{eq:interferometer} only realizes the BAE measurements of $\boldsymbol{q}_{\mathrm{out}}$ with respect to $\boldsymbol{p}_{\mathrm{in}}$. Finally,  by Lemma \ref{lem:coinvariant}, it can be easily verified that the orthogonal and blockwise symplectic matrix
\[
\mathcal{P}=\frac{1}{\sqrt{2}}\left[
\begin{array}{cccc}
1 & 1 & 0 &0\\
0 & 0 & 1 &1\\
1 & -1 & 0 &0\\
0 & 0 & 1 &-1\\
\end{array}
\right]
\]
transforms the system \eqref{eq:interferometer}  to two controllable and observable subsystems which are decoupled from each other, i.e., $G_{co} \boxplus G_{co}$.
\end{example}

\begin{remark}
It is worthwhile to notice that the properties  of system (2) in \cite{TC10} and the system in Fig. 3(b) of   \cite{NY14} can also be checked by using Theorem \ref{thm:BAE}.

\end{remark}



\section{Conclusion} \label{sec:con}
In this paper, we have investigated the structure of quantum linear systems by means of their Kalman canonical form.  In particular,  a new parametrization method has been proposed which generates the quantum Kalman canonical form directly.  Necessary and sufficient conditions for realizing  quantum BAE measurements have also been proposed in terms of these physical parameters.  The {\it system analysis} results presented in this paper may be useful for quantum control engineering, e.g., of opto-mechanical systems.


\textbf{Appendix.}

\textit{Proof of Lemma \ref{lem:H_Gamma}.} Suppose that the system Hamiltonian $\boldsymbol{H}$ in Eq. (\ref{H}) and the coupling operator $\boldsymbol{L}$ in Eq. (\ref{L}) indeed lead to QSDEs in  the
Kalman canonical form (\ref{real_Kalman_ss}). However, in the annihilation-creation operator representation, $\boldsymbol{H}$ and $\boldsymbol{L}$ also lead to  the QSDEs (\ref{eq:sys_a}). As shown in Subsection \ref{sec:zgpg18}, the QSDEs (\ref{eq:sys_a}) and the QSDEs (\ref{real_Kalman_ss}) are related by the coordinate transformations
(\ref{complex-real}). Thus, by Eq. (\ref{complex-real}), we have
\begin{equation*}
\boldsymbol{H}
=
\frac{1}{2}\boldsymbol{x}^{\dag }H\boldsymbol{x}
=
\frac{1}{2}\boldsymbol{\breve{a}}^{\dag }\hat{T}H\hat{T}^{\dag }\boldsymbol{\breve{a}}.
\end{equation*}
Therefore,
\begin{equation} \label{H_real}
H=\hat{T}^{\dag }\Omega \hat{T}.
\end{equation}
Moreover, by Eqs. \eqref{TJT}-\eqref{complex-real}, we get
\begin{align}
\hat{T}^{\dag }J_{n}\hat{T}  = \imath \mathbb{\bar{J}}_{n}.
\label{temp2}
\end{align}
Substituting Eqs.  (\ref{Omega})  and (\ref{temp2}) into Eq. (\ref{H_real}),
together with Lemma \ref{cor:ABC}, we have
\begin{eqnarray*}
H
&=&
\frac{\imath}{2}\hat{T}^{\dag }\left( J_{n}\mathcal{A} - {A}^{\dag}J_{n}\right) \hat{T}
\\
&=&
\frac{1}{2}\left( \bar{A}^{\top}\mathbb{\bar{J}}_{n}-\mathbb{\bar{J}}_{n}\bar{A}\right)
 \\
&=&
\left[
\begin{array}{cc}
0 & -A_{h}^{22} \\
-A_{h}^{22^{\top}} & A_{h}^{12}-B_{h}\mathbb{J}_{m}B_{h}^{\top}/2 \\
0 & A_{12}^{\top}-\mathbb{J}_{n_{1}}B_{co}\mathbb{J}_{m}B_{h}^{\top}/2 \Tstrut\\
0 & A_{13}^{\top}
\end{array}
\right.
\\
&& \hspace{6mm}
\left.
\begin{array}{cc}
0 & 0 \\
A_{12}-B_{h}\mathbb{J}_{m}B_{co}^{\top}\mathbb{J}_{n_{1}}/2 & A_{13}\\
 -\mathbb{J}_{n_{1}}A_{co}+\mathbb{J}_{n_{1}}B_{co}\mathbb{J}_{m}B_{co}^{\top}\mathbb{J}_{n_{1}}/2 & 0 \Tstrut  \\
  0 & -\mathbb{J}_{n_{2}}A_{\bar{c}\bar{o}}
\end{array}
\right],
\end{eqnarray*}
which yields Eq. \eqref{dec12_3}.  On the other hand, from
\begin{equation*}
\left[
\begin{array}{c}
\boldsymbol{L} \\
\boldsymbol{L}^{\#}
\end{array}
\right]
=
\left[
\begin{array}{c}
\Gamma\\
\Gamma^{\#}
\end{array}
\right] \boldsymbol{x}
=
\left[
\begin{array}{c}
\Gamma\\
\Gamma^{\#}
\end{array}
\right] \hat{T}^{\dag }\boldsymbol{\breve{a}}
=
\mathcal{C}\boldsymbol{\breve{a}},
\end{equation*}
we have
\begin{equation}
\left[
\begin{array}{c}
\Gamma\\
\Gamma^{\#}
\end{array}
\right]
=
\mathcal{C}\hat{T}.
\label{gamma2}
\end{equation}
 Moreover, by Eqs. \eqref{complex-real}, \eqref{real_Kalman_sys_A}, and (\ref{gamma2}), we get
\begin{eqnarray*}
\left[
\begin{array}{c}
\Gamma\\
\Gamma^{\#}
\end{array}
\right]
&=&
V_{m}^{\dag }\bar{C} =
\left[
\begin{array}{cc|c|c}
0 & V_{m}^{\dag }C_{h} & V_{m}^{\dag }C_{co} & 0
\end{array}
\right] ,
\end{eqnarray*}
which yields Eq. (\ref{Gamma}).  \hfill $
\square $

{\it Proof of Lemma \ref{thm:ABC_H_Gamma}.} As there are $m$ input fields, we write the coupling operator $\boldsymbol{L}$ as $\boldsymbol{L}
=[
\boldsymbol{L}_{1} \  \cdots  \ \boldsymbol{L}_{m}]^\top$. Given the system Hamiltonian
$\boldsymbol{H}$ and coupling operator $\boldsymbol{L}$, the temporal evolution of a system variable
$\boldsymbol{X}$ is given by, (\cite{GZ00}, \cite{JNP08}, \cite{GJN10}),
\begin{align*}
&d\boldsymbol{X}(t)
\\
=&
 -\imath[\boldsymbol{X}(t),\boldsymbol{H}(t)]dt
 +\frac{1}{2}\sum_{j=1}^{m}\boldsymbol{L}_{j}(t)^{\ast }[\boldsymbol{X}(t), \ \boldsymbol{L}_{j}(t)]dt
\\
&
+\frac{1}{2}\sum_{j=1}^{m}[\boldsymbol{L}_{j}(t)^{\ast },\boldsymbol{X}(t)]\boldsymbol{L}_{j}(t)dt
\\
&
+\sum_{j=1}^{m}d\boldsymbol{B}_{j}(t)^{\ast }[\boldsymbol{X}(t),\ \boldsymbol{L}_{j}(t)]
+\sum_{j=1}^{m}[\boldsymbol{L}_{j}(t)^{\ast },\ \boldsymbol{X}(t)]d\boldsymbol{B}_{j}(t),
\end{align*}
where $\bold{B}_{j}(t)\equiv \int_{0}^t \boldsymbol{b}_{j}(\tau )d\tau $ are quantum Wiener processes ($j=1,\ldots, m$). The above equation
can be re-written in a more compact form as
\begin{align}
&
d\boldsymbol{X}(t)
\nonumber
\\
=
&-\imath[\boldsymbol{X}(t),\boldsymbol{H}(t)]dt
\nonumber
\\
&
-\frac{1}{2}\boldsymbol{L}(t)^{\dag }[\boldsymbol{L}(t),\boldsymbol{X}(t)]dt
+\frac{1}{2}\boldsymbol{L}(t)^{\top}[\boldsymbol{L}(t)^{\#},\boldsymbol{X}(t)]dt
\nonumber
\\
&
-d\boldsymbol{B}(t)^{\dag }[\boldsymbol{L}(t),\boldsymbol{X}(t)]
+d\boldsymbol{B}(t)^{\top}[\boldsymbol{L}(t)^{\#},\boldsymbol{X}(t)]
\nonumber
\\
=&
-\imath[\boldsymbol{X}(t),\boldsymbol{H}(t)]dt
-\frac{1}{2}\boldsymbol{\breve{L}}(t)^{\dag }J_{m}[ \boldsymbol{\breve{L}}(t),\boldsymbol{X}(t)]dt
\nonumber
\\
&
-d\boldsymbol{\breve{B}}(t)^{\dag }J_{m}[ \boldsymbol{\breve{L}}(t),\boldsymbol{X}(t)] .
\label{dX_t}
\end{align}
Informally, Eq. (\ref{dX_t}) can be re-written as
\begin{eqnarray}
\boldsymbol{\dot{X}}(t)
&=&
-\imath[\boldsymbol{X}(t),\boldsymbol{H}(t)]-\frac{1}{2}
\boldsymbol{\breve{L}}(t)^{\dag }J_{m}[ \boldsymbol{\breve{L}}(t),\ \boldsymbol{X}(t)]
\nonumber
\\
 &&
 -\boldsymbol{\breve{b}}(t)^{\dag }J_{m}[\boldsymbol{\breve{L}}(t),\ \boldsymbol{X}(t)] .
 \label{dX_t2}
\end{eqnarray}
It should be noted that Eq. (\ref{dX_t2}) should be understood as (\ref{dX_t}). Using the coordinate transformations (\ref{complex-real}), Eq. (\ref{dX_t2}) becomes
\begin{eqnarray}
\boldsymbol{\dot{X}}(t)
&=&
-\imath[\boldsymbol{X}(t),\boldsymbol{H}(t)]-\frac{1}{2}
\boldsymbol{\breve{L}}(t)^{\dag }J_{m}[ \boldsymbol{\breve{L}}(t),\ \boldsymbol{X}(t)]
\nonumber
\\
&&
-\boldsymbol{u}(t)^{\top}V_{m}J_{m}[ \boldsymbol{\breve{L}}(t),\ \boldsymbol{X}(t)] .
\label{dX_t3}
\end{eqnarray}
Substituting the elements of $\boldsymbol{x}$ into Eq. (\ref{dX_t3}) and transposing both sides of the resulting equation, we have
\begin{eqnarray}
\boldsymbol{\dot{x}}(t)^\top
&=&
-\imath[\boldsymbol{x}(t)^\top, \ \boldsymbol{H}(t)]-\frac{1}{2}
\boldsymbol{\breve{L}}(t)^{\dag }J_{m}[ \boldsymbol{\breve{L}}(t),\ \boldsymbol{x}(t)^\top ]
\nonumber
\\
&&
-\boldsymbol{u}(t)^{\top}V_{m}J_{m}[ \boldsymbol{\breve{L}}(t), \ \boldsymbol{x}(t)^\top] .  \label{dX_t3b}
\end{eqnarray}
After system-field interaction, the output fields
\begin{equation*}
\boldsymbol{\breve{b}}_{\mathrm{out}}(t)
=
\boldsymbol{\breve{L}}(t)+\boldsymbol{\breve{b}}(t),
\end{equation*}
 are generated, which, by the coordinate transformations (\ref{complex-real}),  in the real quadrature operator representation are
\begin{equation}\label{y}
\boldsymbol{y}(t)=V_{m}\boldsymbol{\breve{L}}(t)+\boldsymbol{u}(t).
\end{equation}
Given the matrix $H$ in Eq. (\ref{R}), the Hamiltonian $\boldsymbol{H}$  in Eq. (\ref{H}) can be re-written as
\begin{eqnarray}
\boldsymbol{H}
&=&
\frac{1}{2}\boldsymbol{q}_{h}^{\top}H_{h}^{12}\boldsymbol{p}_{h}+\frac{1}{2}
\boldsymbol{p}_{h}^{\top}H_{h}^{12^{\top}}\boldsymbol{q}_{h}+\frac{1}{2}
\boldsymbol{p}_{h}^{\top}H_{h}^{22}\boldsymbol{p}_{h}
 \nonumber
 \\
&&
+\frac{1}{2}\boldsymbol{p}_{h}^{\top}H_{12}\boldsymbol{x}_{co} +\frac{1}{2}\boldsymbol{x}_{co}^{\top}H_{12}^{\top}\boldsymbol{p}_{h}
\nonumber
 \\
&&
+\frac{1}{2}
\boldsymbol{x}_{co}^{\top}H_{co}\boldsymbol{x}_{co}  +\frac{1}{2}\boldsymbol{x}_{\bar{c}\bar{o}}^{\top}H_{\bar{c}\bar{o}}\boldsymbol{x}_{\bar{c}\bar{o}}
\nonumber
 \\
&&
+\frac{1}{2}
\boldsymbol{p}_{h}^{\top}H_{13}\boldsymbol{x}_{\bar{c}\bar{o}}+\frac{1}{2}\boldsymbol{x}_{\bar{c}\bar{o}}^{\top}H_{13}^{\top}\boldsymbol{p}
_{h}.
 \label{H_2}
\end{eqnarray}
After standard, although tedious calculation, one can obtain
\begin{equation}\label{x_H}
-\imath[\boldsymbol{x},\boldsymbol{H}]
=
\left[
\begin{array}{cccc}
H_{h}^{12^{\top}} & H_{h}^{22} & H_{12} & H_{13} \\
0 & -H_{h}^{12} & 0 & 0 \Tstrut \\
0 & \mathbb{J}_{n_{1}}H_{12}^{\top} & \mathbb{J}_{n_{1}}H_{co} & 0 \Tstrut  \\
0 & \mathbb{J}_{n_{2}}H_{13}^{\top} & 0 & \mathbb{J}_{n_{2}}H_{\bar{c}\bar{o}} \Tstrut
\end{array}
\right] \boldsymbol{x},
\end{equation}
and
\begin{equation}\label{L_x}
[ \boldsymbol{\breve{L}}(t),\boldsymbol{x}(t)^{\top}]
=
-\imath\left[
\Gamma_{h} \  0 \  -\Gamma_{co}\mathbb{J}_{n_{1}} \  0
\right] .
\end{equation}
Substituting Eqs. (\ref{x_H})-(\ref{L_x}) into (\ref{dX_t3b}) we get
\begin{align*}
\boldsymbol{\dot{x}}(t)
=&\;
(\boldsymbol{\dot{x}}(t)^\top)^\dag
\nonumber
\\
=&\;
\left[
\begin{array}{cccc}
H_{h}^{12^{\top}} & H_{h}^{22} & H_{12} & H_{13} \\
0 & -H_{h}^{12} & 0 & 0 \Tstrut \\
0 & \mathbb{J}_{n_{1}}H_{12}^{\top} & \mathbb{J}_{n_{1}}H_{co} & 0 \Tstrut \\
0 & \mathbb{J}_{n_{2}}H_{13}^{\top} & 0 & \mathbb{J}_{n_{2}}H_{\bar{c}\bar{o}} \Tstrut
\end{array}
\right] \boldsymbol{x}(t)
 \\
&
\;
-\frac{\imath}{2}\left[
\begin{array}{cccc}
0 & \Gamma_{h}^{\dag }J_{m}\Gamma_{h} & \Gamma_{h}^{\dag
}J_{m}\Gamma_{co} & 0 \\
0 & 0 & 0 & 0 \\
0 & \mathbb{J}_{n_{1}}\Gamma_{co}^{\dag }J_{m}\Gamma_{h} & \mathbb{J}
_{n_{1}}\Gamma_{co}^{\dag }J_{m}\Gamma_{co} & 0  \\
0 & 0 & 0 & 0
\end{array}
\right] \boldsymbol{x}(t)
\\
&\;
+\left[
\begin{array}{c}
\Gamma_{h}^{\dag }V_{m}^{\dag }\mathbb{J}_{m} \\
0 \\
\mathbb{J}_{n_{1}}\Gamma_{co}^{\dag }V_{m}^{\dag }\mathbb{J}_{m} \\
0
\end{array}
\right] \boldsymbol{u}(t),
\end{align*}
from which Eq.  \eqref{dec12_1} follows. Finally, by Eq. (\ref{Gamma}),
\begin{equation}\label{eq:dec26}
V_{m}\boldsymbol{\breve{L}}
=
V_{m}(\Gamma_{h}\boldsymbol{p}_{h}+\Gamma_{co}\boldsymbol{x}_{co}).
\end{equation}
Substituting  Eq. (\ref{eq:dec26}) into Eq. (\ref{y}) yields
 Eq. \eqref{dec12_2}. \hfill $\Box$

{\it Proof of Lemma \ref{lem:equiv_1}.} This result is a consequence of Lemmas \ref{cor:ABC} and  \ref{thm:ABC_H_Gamma}.  Notice
$
\left( A_{h}^{11},B_{h}\right)
=
( -A_{h}^{22^{\top}},\Gamma_{h}^{\dag}V_{m}^{\dag }\mathbb{J}_{m})
=
( H_{h}^{12^{\top}},\Gamma_{h}^{\dag }V_{m}^{\dag }\mathbb{J}_{m}).
$
As a result,
$( A_{h}^{11},B_{h}) $ is controllable $\Longleftrightarrow $ $
( H_{h}^{12^{\top}},\Gamma_{h}^{\dag }V_{m}^{\dag }\mathbb{J}_{m}) $ is controllable
$\Longleftrightarrow $ $( H_{h}^{12},\mathbb{J}_{m}V_{m}\Gamma_{h}) $ is observable
$\Longleftrightarrow $
$(( H_{h}^{12},\Gamma_{h})$ is observable. This establishes the equivalence between (i) and (iii).  On the other hand, because
$( A_{h}^{22},C_{h}) =( -H_{h}^{12},V_{m}\Gamma_{h})$,
$( A_{h}^{22},C_{h}) $ is observable $\Longleftrightarrow $
$( -H_{h}^{12},V_{m}\Gamma_{h}) $ is observable $\Longleftrightarrow $
$( H_{h}^{12},\Gamma_{h}) $ is observable. This establishes the equivalence between (ii) and (iii). \hfill $\square $

{\it Proof of Lemma \ref{lem:equiv_2}.} This result can be proved in a similar way as in the proof of  Lemma \ref{lem:equiv_1}. Notice that
$
( A_{co},B_{co})
=
( \mathbb{J}_{n_{1}}H_{co}-\imath\mathbb{J}_{n_{1}}\Gamma_{co}^{\dag }J_{m}\Gamma_{co}/2,
\mathbb{J}_{n_{1}}\Gamma_{co}^{\dag }V_{m}^{\dag }\mathbb{J}_{m}).
$
Hence,
$( A_{co},B_{co}) $ is controllable $\Longleftrightarrow $
$( \mathbb{J}_{n_{1}}H_{co}-\imath\mathbb{J}_{n_{1}}\Gamma_{co}^{\dag}J_{m}\Gamma_{co}/2,
\mathbb{J}_{n_{1}}\Gamma_{co}^{\dag }V_{m}^{\dag }\mathbb{J}_{m}) $ is controllable $\Longleftrightarrow $ $(\mathbb{J}_{n_{1}}H_{co},\mathbb{J}_{n_{1}}\Gamma_{co}^{\dag }) $
is controllable. The last statement follows  since if
$x^{\dag }(\mathbb{J}_{n_{1}}H_{co}-\imath\mathbb{J}_{n_{1}}\Gamma_{co}^{\dag }J_{m}\Gamma_{co}/2)
=\lambda x^{\dag }$
and
$x^{\dag }\mathbb{J}_{n_{1}}\Gamma_{co}^{\dag }V_{m}^{\dag }\mathbb{J}_{m}
=0$, then
$x^{\dag }\mathbb{J}_{n_{1}}\Gamma_{co}^{\dag}=0$. As a result,
$x^{\dag }\mathbb{J}_{n_{1}}H_{co}=\lambda x^{\dag} $. On the other hand, if
$x^{\dag }\mathbb{J}_{n_{1}}H_{co}
=\lambda x^{\dag }$ and $x^{\dag }\mathbb{J}_{n_{1}}\Gamma_{co}^{\dag }=0$,
then
$x^{\dag }\mathbb{J}_{n_{1}}\Gamma_{co}^{\dag }V_{m}^{\dag }\mathbb{J}_{m}=0$ and
$x^{\dag }(\mathbb{J}_{n_{1}}H_{co}-\imath\mathbb{J}_{n_{1}}\Gamma_{co}^{\dag }J_{m}\Gamma_{co}/2)=x^{\dag }\mathbb{J}_{n_{1}}H_{co}=\lambda x^{\dag }$. Now
$(\mathbb{J}_{n_{1}}H_{co},\mathbb{J}_{n_{1}}\Gamma_{co}^{\dag }) $ is controllable  $\Longleftrightarrow $
$( H_{co}\mathbb{J}_{n_{1}},\Gamma_{co}^{\dag }) $ is controllable
$\Longleftrightarrow ( \mathbb{J}_{n_{1}}H_{co},\Gamma_{co}) $ is
observable. This establishes the equivalence between (i) and (iii). On the other hand, notice that
$
( A_{co},C_{co}) =( \mathbb{J}_{n_{1}}H_{co}-\imath\mathbb{J}
_{n_{1}}\Gamma_{co}^{\dag }J_{m}\Gamma_{co}/2,V_{m}\Gamma_{co}).
$
Hence,
$( A_{co},C_{co}) $ is observable $\Longleftrightarrow $ $(
\mathbb{J}_{n_{1}}H_{co}-\imath\mathbb{J}_{n_{1}}\Gamma_{co}^{\dag
}J_{m}\Gamma_{co}/2,V_{m}\Gamma_{co}) $ is observable $
\Longleftrightarrow $ $( \mathbb{J}_{n_{1}}H_{co},V_{m}\Gamma
_{co}) $ is observable. The last statement holds  since if $( \mathbb{J}_{n_{1}}H_{co}-i%
\mathbb{J}_{n_{1}}\Gamma_{co}^{\dag }J_{m}\Gamma_{co}/2)
x=\lambda x$, and $V_{m}\Gamma_{co}x=0$, then $\Gamma_{co}x=0$ and $
\mathbb{J}_{n_{1}}H_{co}x=\lambda x$. On the other hand, if $\Gamma_{co}x=0$
and $\mathbb{J}_{n_{1}}H_{co}x=\lambda x$, then $V_{m}\Gamma_{co}x=0$ and $
( \mathbb{J}_{n_{1}}H_{co}-\imath\mathbb{J}_{n_{1}}\Gamma_{co}^{\dag
}J_{m}\Gamma_{co}/2) x=\mathbb{J}_{n_{1}}H_{co}x=\lambda x$. Now $( \mathbb{J}_{n_{1}}H_{co},V_{m}\Gamma
_{co}) $ is observable  $
\Longleftrightarrow $ $( \mathbb{J}_{n_{1}}H_{co},\Gamma_{co}) $
is observable. This establishes the equivalence between (ii) and (iii). \hfill $\square $

{\it Proof of Lemma \ref{lem:equiv_mar25}.}  According to \cite[Proposition 1]{GZ15}, the controllability and observability of  the system (\ref{real_Kalman_ss}) are equivalent.  Clearly, the system (\ref{real_Kalman_ss})  is controllable if and only if the following subsystem
\begin{eqnarray}\label{sys:c}
 \left[\begin{array}{c}
\boldsymbol{\dot{x}}_{co}(t)\\
\boldsymbol{\dot{q}}_h(t)
\end{array}
\right] =
\left[
\begin{array}{cc}
A_{co} & 0\\
A_{12}  & A_h^{11}
\end{array}
\right] \left[\begin{array}{c}
\boldsymbol{x}_{co}(t)\\
\boldsymbol{q}_h(t)
\end{array}
\right]
+ \left[\begin{array}{c}
B_{co}\\
B_h
\end{array}
\right]\boldsymbol{u}(t)
\end{eqnarray}
is controllable.  By Lemma \ref{thm:ABC_H_Gamma}, the system (\ref{sys:c}) is controllable if and only if $\left( \left[
\begin{array}{cc}
\mathbb{J}_{n_{1}}H_{co} & 0\\
H_{12}  & H_h^{12^\top}
\end{array}
\right],  \left[\begin{array}{c}
\mathbb{J}_{n_{1}}\Gamma_{co}^\dag \\
\Gamma_h^\dag
\end{array}
\right] \right)$ is  controllable. On the other hand,  the system (\ref{real_Kalman_ss})  is observable if and only if the following subsystem
\begin{equation}\label{sys:o}
\begin{split}
 \left[\begin{array}{c}
\boldsymbol{\dot{x}}_{co}(t)\\
\boldsymbol{\dot{p}}_h(t)
\end{array}
\right] =&
\left[
\begin{array}{cc}
A_{co} & A_{21}\\
0  &  A_h^{22}
\end{array}
\right] \left[\begin{array}{c}
\boldsymbol{x}_{co}(t)\\
\boldsymbol{p}_h(t)
\end{array}
\right]
+ \left[\begin{array}{c}
B_{co}\\
0
\end{array}
\right]\boldsymbol{u}(t),
\\
\boldsymbol{y}(t) =&  \left[\begin{array}{cc}
C_{co} & C_h
\end{array}
\right]  \left[\begin{array}{c}
\boldsymbol{x}_{co}(t)\\
\boldsymbol{p}_h(t)
\end{array}
\right]+\boldsymbol{u}(t)
\end{split}
\end{equation}
is observable. However, by Lemma \ref{thm:ABC_H_Gamma}, the system (\ref{sys:o}) is observable if and only if  $\left( \left[
\begin{array}{cc}
\mathbb{J}_{n_{1}}H_{co} & \mathbb{J}_{n_{1}}H_{12}^\top\\
0  & -H_h^{12}
\end{array}
\right],  \left[\begin{array}{cc}
\Gamma_{co} &
\Gamma_h
\end{array}
\right] \right)$ is observable. \hfill $\Box $

{\it Proof of Lemma \ref{lem:coinvariant}.} Define $\boldsymbol{H}_{co1}\triangleq\frac{1}{2}\boldsymbol{x}_{co1}^TH_{co1}\boldsymbol{x}_{co1}$, $\boldsymbol{H}_{co2}\triangleq\frac{1}{2}\boldsymbol{x}_{co2}^TH_{co2}\boldsymbol{x}_{co2}$, and $\boldsymbol{H}_{coh}\triangleq\frac{1}{2}\boldsymbol{p}_{h}^TH_{121}\boldsymbol{x}_{co1}+\frac{1}{2}\boldsymbol{x}_{co1}^TH_{121}^T\boldsymbol{p}_{h}$. Since $H_{co}$ and $H_{12}$ are in the form \eqref{eq:cohgamma}, the system Hamiltonian $\boldsymbol{H}_{co}$ for the ``$co$'' subsystem can be rewritten as follows
\begin{equation}\label{eq:cohamiltonian}
\boldsymbol{H}_{co}=\boldsymbol{H}_{co1}+\boldsymbol{H}_{co2}+\boldsymbol{H}_{coh}.
\end{equation}
On the other hand, it is worth mentioning that $\boldsymbol{L}$ is a column vector whose elements represent the coupling of each field with the quantum system. This means that swapping the elements in $\boldsymbol{L}$ does not change the coupling relationship between the fields and the quantum system. Since condition (B2) holds, we can re-arrange the elements in $\boldsymbol{L}$ to transform it into the following form
\begin{equation}\label{eq:cocoupling}
\left[\begin{array}{c}\boldsymbol{L}_{co1}\\ \boldsymbol{L}_{co2}\end{array}\right].
\end{equation}
where $\left[\begin{array}{c}\boldsymbol{L}_{co2}\\ \boldsymbol{L}_{co2}^\#\end{array}\right]=\hat{\Gamma}_{co2}\boldsymbol{x}_{co2}$. By Definition \ref{defn:invariant}, Eqs. \eqref{eq:cohamiltonian} and \eqref{eq:cocoupling} imply that the subsystem $G_{co}$ in  \eqref{eq:coinvariant} is an invariant subsystem. In terms of the form \eqref{eq:cohgamma}, it follows from Lemma \ref{lem:equiv_2} that $G_{co}$ is both controllable and observable. \hfill $\Box$

{\it Proof of Theorem \ref{thm:BAE}.} From Eqs. \eqref{Gamma_co0} and (\ref{dec12_2}), we have%
\begin{equation}\label{key}
\begin{split}
  C_{co,q}=&\ \sqrt{2}\left[
\begin{array}{cc}
\mathrm{Re}\left( \Gamma_{co,q}\right) & \mathrm{Re}\left( \Gamma_{co,p}\right)
\end{array}
\right],\\
C_{co,p}=&\ \sqrt{2}\left[
\begin{array}{cc}
\mathrm{Im}\left( \Gamma_{co,q}\right) & \mathrm{Im}\left( \Gamma_{co,p}\right)
\end{array}
\right].
\end{split}
\end{equation}
By Eq. (\ref{pr1b_b}), the following can be obtained
\begin{eqnarray}
B_{co,q} = -\mathbb{J}_{n_{1}}C_{co,p}^{\top},  \ \ \
B_{co,p} =\mathbb{J}_{n_{1}}C_{co,q}^{\top}.
 \label{B_coq}
\end{eqnarray}
Moreover, by Eq. (\ref{dec12_3}), we have
\begin{equation}\label{JnA_co}
\begin{split}
A_{co}=&\ \mathbb{J}_{n_{1}}H_{co}+B_{co}\mathbb{J}_{m}B_{co}^{\top}\mathbb{J}
_{n_{1}}/2 \\
=&\ \mathbb{J}_{n_{1}}H_{co}-\mathbb{J}_{n_{1}}C_{co,p}^{\top}C_{co,q}/2+\mathbb{%
J}_{n_{1}}C_{co,q}^{\top}C_{co,p}/2.
\end{split}
\end{equation}

(i) According to Eq. \eqref{tf:pq}, Eq. \eqref{tf:pq0} is equivalent to
\begin{equation}\label{eq:baeabc}
  C_{co,q}A_{co}^kB_{co,p}=0,\ \ k=0,1,\cdots.
\end{equation}
Moreover, by Eqs. \eqref{key} and \eqref{B_coq}, Eq. \eqref{temp9} is  equivalent to
\begin{equation}\label{eq:baehgamma}
  C_{co,q}(\mathbb{J}_{n_{1}}H_{co})^kB_{co,p}=0,\   \ k=0,1,\cdots.
\end{equation}
Thus, it suffices to establish the equivalence between Eqs. \eqref{eq:baeabc} and \eqref{eq:baehgamma}.

Firstly, we show that Eq. \eqref{eq:baehgamma} implies Eq. \eqref{eq:baeabc}. We do this by induction. Suppose Eq. \eqref{eq:baehgamma} holds.  Then
\begin{equation}\label{eq:CB_feb27}
C_{co,q} B_{co,p}=0.
\end{equation}
 Assume that
\begin{equation}\label{eq:CAB_feb28a}
 C_{co,q}A_{co}^l B_{co,p}=0, \ \  \forall l\leq k-1.
\end{equation}
   By Eqs. \eqref{B_coq} and \eqref{JnA_co}, direct matrix manipulations yield
\begin{equation*}\label{eq:baek}
\begin{split}
  &\ C_{co,q}A_{co}^kB_{co,p}
  \\
  =&\ C_{co,q}\mathbb{J}_{n_{1}}H_{co}A_{co}^{k-1}B_{co,p}
  \\
  &\ -C_{co,q}\mathbb{J}_{n_{1}}C_{co,p}^{\top}\boxed{C_{co,q}A_{co}^{k-1}B_{co,p}}/2
  \\
  &\ +\boxed{C_{co,q}B_{co,p}}C_{co,p}A_{co}^{k-1}B_{co,p}/2
\\
=&\ C_{co,q}(\mathbb{J}_{n_{1}}H_{co})A_{co}^{k-1}B_{co,p}
\\
=&\ \cdots
\\
=&\ C_{co,q}(\mathbb{J}_{n_{1}}H_{co})^k B_{co,p}
\\
=&\ 0,
\end{split}
\end{equation*}
where the two terms in the boxes above are both equal to  zero due to Eqs. \eqref{eq:CB_feb27} and \eqref{eq:CAB_feb28a}. Therefore, by mathematical induction, Eq. \eqref{eq:baeabc} holds.

Secondly,  assume that Eq. \eqref{eq:baeabc} holds.  Clearly. Eq. \eqref{eq:CB_feb27} holds.  Assume that
\begin{equation}\label{eq:CAB_feb28b}
C_{co,q}(\mathbb{J}_{n_{1}}H_{co})^l B_{co,p}=0, \ \ \forall l \leq k-1.
\end{equation}
 By Eqs. \eqref{B_coq} and \eqref{JnA_co},
\begin{eqnarray*}
&&C_{co,q}(\mathbb{J}_{n_{1}}H_{co})^{k}B_{co,p} \\
&=&C_{co,q}A_{co}(\mathbb{J}_{n_{1}}H_{co})^{k-1}B_{co,p} \\
&&+C_{co,q}\mathbb{J}_{n_{1}}C_{co,p}^{\top }\boxed{C_{co,q}(\mathbb{J}%
_{n_{1}}H_{co})^{k-1}B_{co,p}}/2 \\
&&-\boxed{C_{co,q}B_{co,p}}C_{co,p}(\mathbb{J}%
_{n_{1}}H_{co})^{k-1}B_{co,p}/2 \\
&=&C_{co,q}A_{co}(\mathbb{J}_{n_{1}}H_{co})^{k-1}B_{co,p} \\
&=&\cdots  \\
&=&C_{co,q}A_{co}^{k}B_{co,p}
\\
&=&
0,
\end{eqnarray*}
where the two terms in the boxes above are both equal to zero due to Eqs. \eqref{eq:CB_feb27} and \eqref{eq:CAB_feb28b}. Thus, by mathematical induction, Eq. \eqref{eq:baehgamma} holds.  Thus the equivalence between Eqs. \eqref{eq:baeabc} and \eqref{eq:baehgamma} has been established.

(ii) The proof  follows in a similar way as that of (i), and thus is omitted. \hfill $\Box$

{\it Proof of Corollary \ref{cor:BAE}.} Let $H$ be as  in Eq. (\ref{H_co}). Then $\mathbb{J}_{n }H=J_{n }$ and%
\begin{eqnarray}
&&\mathrm{rank}\left( \left[
\begin{array}{c}
\breve{\Gamma} \\
\breve{\Gamma}\mathbb{J}_{n }H \\
\vdots  \\
\breve{\Gamma}(\mathbb{J}_{n }H)^{2n -1}
\end{array}
\right] \right)
=
\mathrm{rank}\left( \left[
\begin{array}{c}
\breve{\Gamma} \\
\breve{\Gamma}J_{n }
\end{array}
\right] \right) .
\label{rank_Gamma_H}
\end{eqnarray}
Therefore, the observability of $( \mathbb{J}_{n }H,\breve{\Gamma}) $ is equivalent to Eq. (\ref{rank Gamma}). In a similar way, by
Lemma \ref{lem:equiv_2},  the controllability of the system is
also equivalent to Eq. (\ref{rank Gamma}).

(i) By Eq. (\ref{H_co}),  it can be seen that Eq. (\ref{temp9}) in Theorem \ref{thm:BAE} is equivalent to%
\begin{align}
\mathrm{Re}(\Gamma)(\mathbb{J}_{n }H)^{k}\mathbb{J}_{n }\mathrm{Re}(\Gamma^{\top})=&\;
\mathrm{Re}(\Gamma) \left[
\begin{array}{cc}
I_{n_1} & 0\\
0            & -I_{n_1}
\end{array}
\right]^{k}\mathbb{J}_{n }\mathrm{Re}(\Gamma^{\top})
\nonumber
\\
 =&\;
 0,\ \ k=0,1,\ldots .  \label{temp8}
\end{align}
However, Eq. (\ref{temp8}) is equivalent to
\[
\mathrm{Re}\left( \Gamma_{q}\right)\mathrm{Re}\left( \Gamma_{p} ^{\top}\right)
-\mathrm{Re}\left( \Gamma_{p}\right)\mathrm{Re}\left( \Gamma_{q}^{\top}\right)= 0,
\]
and
\[
\mathrm{Re}\left( \Gamma_{q}\right)\mathrm{Re}\left( \Gamma_{p}^{\top}\right)
+
\mathrm{Re}\left( \Gamma_{p}\right)\mathrm{Re}\left( \Gamma_{q}^{\top}\right)=0,
\]
which are equivalent to Eq. (\ref{perp_1a}).

(ii) In a similar way, it can be shown that Eq. (\ref{temp9b}) in Theorem \ref{thm:BAE} is equivalent to
\[
\mathrm{Im}\left( \Gamma_{q}\right)\mathrm{Im}(\Gamma_{p}^{\top})
-\mathrm{Im}\left( \Gamma_{p}\right)\mathrm{Im}( \Gamma_{q}^{\top})= 0,
\]
and
\[
\mathrm{Im}\left( \Gamma_{q}\right)\mathrm{Im}\left(\Gamma_{p}^{\top}\right)
+\mathrm{Im}\left( \Gamma_{p}\right)\mathrm{Im}\left( \Gamma_{q}^{\top}\right)=0,
\]
which are equivalent to Eq. (\ref{perp_2a}).  \hfill $\Box $

\bibliographystyle{IFAC}

\end{document}